\documentclass[%
 reprint,
 amsmath,amssymb,
 aps,
]{revtex4-2}

\usepackage{graphicx}
\usepackage{dcolumn}
\usepackage{bm}
\usepackage{subfigure}
\usepackage{mathtools}

\begin{document}

\preprint{APS/123-QED}

\title{Cosmological Simulation with Population III Stellar Feedback and Metal Enrichment I: Model Description And Convergence Test}

\author{Bocheng Zhu}
\email{bochengzhu@outlook.com}
 \affiliation{Institute of Astrophysics, School of Physics, Zhengzhou University, China}
\author{Liang Gao}%
 \email{lgao@bao.ac.cn}
\affiliation{Institute of Astrophysics, School of Physics, Zhengzhou University, China}%
\affiliation{Institute for Frontiers in Astronomy and Astrophysics, Beijing Normal University, Beijing 102206, China
}

\date{\today}

\begin{abstract}
We present SPARK (Simulation of Pop III in Arepo with Radiation and Kinetic feedback) framework, a new Pop III + Pop II subgrid framework implemented in the moving-mesh code {\sc arepo}, designed to study the impact of Pop III feedback on star formation in the early universe. The framework combines primordial non-equilibrium chemistry, metal-line cooling, IMF-sampled stellar evolution with SN feedback, and approximate Lyman-Werner (LW) and ionizing radiation transport. We run a suite of $1c{\rm Mpc}/h$ box simulations with different initial conditions and resolutions from $z=127$ to $z=10$. The highest gas mass and spatial resolution in the fiducial simulation reach $\sim10\,{\rm M_{\odot}}$ and $\sim4\,{\rm pc}$, respectively. The model successfully reproduces the Pop II star formation rate density (SFRD) consistent with previous theoretical works across all initial conditions, with minor variation mainly driven by local halo interactions and LW irradiation. We find that the volume filling factor of metal-enriched gas converges to $\sim0.5-2\%$ at $z=10$, with scatter driven by stochastic and mixing model variations. Convergence is achieved once subhalos with $M_{\rm subhalo}\gtrsim 10^{6.5}\,{\rm M_{\odot}}$  are resolved, and the total stellar mass at $z=10$ is largely insensitive to initial conditions or the resolution considered in this work. A fiducial simulation requires $\sim 10^4$ CPU hours, making the framework computationally tractable for larger box simulations and enabling future large parameter studies of stellar physics or environment effects such as Pop III IMF variations, X-ray radiation, or the streaming velocity at high redshift.
\end{abstract}

\maketitle


\section{Introduction}

The first generation of stars, known as Population III (hereafter Pop III) stars, plays a key role in shaping the initial thermal, chemical, and radiative conditions of the early Universe \citep{bromm04, bromm13, Klessen23}. Forming from pristine, metal-free gas in minihalos at high redshift, these stars provide the first ionizing photons that contribute to cosmic reionization \citep{barkana01}. Furthermore, through supernova (SN) explosions, Pop III stars enrich the pristine intergalactic medium (IGM) with the first heavy elements, fundamentally altering the cooling properties of the gas and regulating the formation of the subsequent metal-enriched Population II (hereafter Pop II) stars and the first galaxies \citep{greif07, wise08, greif10, ritter12, wise12, jeon14, chiaki19, brauer2025, storck25}. Understanding how Pop III stars form and impact their environment in the early universe is therefore essential for understanding the cosmic history.

Despite their crucial importance, direct observation of individual Pop III stars remains elusive due to their faintness and extreme distances. One promising avenue is stellar archaeology, i.e., the observation of metal-poor stars in the Milky Way to seek relics of Pop III stars  \citep{beers05,frebel15}. Massive Pop III stars are predicted to end their lives as gamma-ray bursts \citep{bromm06} or pair-instability supernovae (PISN) \citep{heger02}. These energetic events enrich the surrounding medium, leaving distinctive fingerprints on the second-generation stars formed from the ejecta \citep{karlsson13, hartwig23} and in the spectra of high-redshift GRBs \citep{inoue07,ma15}. Potential PISN signatures have been identified in the Milky Way halo \citep{aoki14,xing23}, and in nearby dwarfs \citep{skuladottir21}.

Complementary to these local archaeological efforts, the recent launch of the James Webb Space Telescope (JWST) has revolutionized our study of the high-redshift Universe. JWST observations have begun to uncover galaxy candidates at $z\gtrsim10$ \citep{castellano22,curtislake23, robertson23, donnan2023}, with a potential Pop III signature found via lensing \citep{maiolino24}. While these observations provide unprecedented constraints, disentangling the signatures of Pop III stars from early Pop II populations is challenging.

On the other hand, indirect constraints from large-scale observables offer a complementary avenue to probe the first stars. Theoretically, the imprint of Pop III stars on the cosmic background radiation has been studied extensively over the past decades. Several works using semi-analytic models \citep{visbal15,mebane18} and semi-numerical codes \citep[e.g., 21cmFAST;][]{mesinger11} have established how the UV and X-ray backgrounds from the first stars likely shaped the thermal and ionization history of the IGM. This field has seen a surge of renewed interest recently, driven by the arrival of observational data. New upper limits on the 21-cm power spectrum from experiments like HERA \citep{hera1} and constraints on the global signal from SARAS 3 \citep{singh22} have begun to place the first stringent constraints on the Pop III parameter space \citep{pochinda24}.

To ground these statistical inferences in robust physics, we also need to understand how variations in Pop III star properties affect the environment and how they, in turn, regulate subsequent star formation. This requires a modeling approach that connects micro-scale stellar physics and macro-scale cosmological simulations. However, modeling Pop III stars in cosmological simulations is challenging. Since Pop III stars form from molecular hydrogen (${\rm H_2}$) clouds, capturing this process requires solving the non-equilibrium chemistry of the primordial gas \citep{abel97, galli98, glover08}. The evolution of Pop III stars is also different from metal-enriched stars in the local universe because of their extremely low metallicity \citep{marigo01, heger02, murphy21}. Furthermore, the initial mass function (IMF) of Pop III stars remains highly uncertain \citep{hirano14, stacy16, jaura22}. Finally, since Pop III stars form in low-mass minihalos, high resolution to resolve them is required for the cosmological simulations \citep{bromm11, greif11}.

Over the past decades, extensive efforts have been dedicated to simulating Pop III star formation and feedback within a cosmological context. Pioneering zoom-in simulations focused on the formation of the first stars in individual minihalos \citep{gao07, yoshida08, greif11, stacy12}. To capture the statistical properties of early galaxies, subsequent studies extended these methods to larger volumes. Notable examples include the Renaissance Simulations \citep{wise12, oshea15, xu16}, as well as other campaigns \citep[e.g.,][]{jaacks18, liu20, sarmento22}. More recently, state-of-the-art simulations have incorporated increasingly detailed physical processes such as on-the-fly radiative transfer (RT), non-equilibrium dust evolution and explicit Pop III modeling, to study the epoch of reionization. Related projects in this category include AEOS \citep{brauer2025}, and MEGATRON \citep{katz25, storck25}. Complementarily, the properties of Pop III stars and their transition to Pop II have been analyzed in high-resolution cosmological zoom simulations \citep[e.g.,][]{zier25}.

These simulations have successfully captured the complex interplay between Pop III star formation, feedback mechanisms, and larger-scale cosmological evolution, providing invaluable insights into how these early stars influence the formation of galaxies. However, despite their remarkable success, these high-resolution cosmological (zoom) simulations are computationally expensive. This high cost limits the feasibility of conducting large parameter space studies, making it challenging to explore how the full range of uncertainties in Pop III models impacts galaxy formation at early epochs. 

In this paper, we present SPARK (Simulation of Pop III in Arepo with Radiation and Kinetic feedback) framework, a new Pop III + Pop II subgrid framework implemented in the moving-mesh code {\sc arepo} \citep{Springel2010}. Our model integrates a primordial non-equilibrium chemical network, metal-line cooling, and a stochastic IMF sampling scheme that explicitly accounts for SN feedback. Furthermore, we incorporate an approximate radiative transfer method to model Lyman-Werner (LW) and ionizing radiation feedback efficiently. As the first paper in a series, our primary goal is to establish a robust framework that is computationally tractable, enabling future extensive investigations into how physical uncertainties related to Pop III stars shape the properties of the early Universe and the formation of the first galaxies.

The paper is organized as follows. We describe the models in detail in Section \ref{sec:method}, including the modeling of star formation, stellar feedback from Pop III and Pop II stars, non-equilibrium chemistry and gas cooling/heating, metal yields, the refinement strategy, and the simulation setup. In Section \ref{sec:dic}, we introduce the main results, including the time evolution, the star formation history,
and the gas properties of the simulations with fiducial resolutions but different initial conditions. In Section \ref{sec:rd}, we investigate the resolution convergence of our model. Finally, we summarize in Section \ref{sec:summary}.

\section{\label{sec:method}Methods}

In this section, we introduce the subgrid models and numerical implementations for star formation, stellar feedback, non-equilibrium primordial chemistry, and radiative cooling within our Pop III cosmological simulation framework. In Section \ref{sec:sf}, we describe the star formation criteria for both Pop III and Pop II populations. The stellar feedback models, including supernova feedback from both populations and stellar winds from massive Pop II stars, are introduced in Section \ref{sec:sfb}. The radiation feedback and our approximate radiative transfer method are presented in Section \ref{sec:rfb}. The scalar turbulent metal mixing model is described in Section \ref{sec:mixing}. The non-equilibrium chemical network and the implementation of radiative cooling are detailed in Section \ref{sec:clchem}. The mesh construction and refinement strategy is described in Section \ref{sec:refinement}. Finally, the specific simulation setup for this work is presented in Section \ref{sec:simset}.

This framework is implemented in \textsc{Arepo}, an $N$-body and moving-mesh magnetohydrodynamical code \citep{Springel2010, pakmor2016improving,Weinberger2020}. Due to its quasi-Lagrangian nature, \textsc{Arepo} naturally provides adaptive resolution and maintains Galilean invariance for fluid dynamics.

\subsection{Star formation}\label{sec:sf}

\subsubsection{Pop III Star formation}\label{sec:pop3sf}

The star formation criteria for the Pop III stars are based on the model implemented in \citet{wise12} with modified density threshold. A gas cell is eligible to form a Pop III star particle if it satisfies the following conditions:

\begin{itemize}
  \item The gas flow is convergent, i.e., $\nabla\cdot \mathbf{v}_{\rm gas} < 0$;
  \item The gas number density $n_{\rm gas}$ exceeds a threshold $n_{\rm th}$ of $\max(1\, {\rm cm^{-3}}, 0.5\, n_{\rm Jeans})$;
  \item The $\rm H_2$ fraction of the gas is higher than $5\times10^{-4}$;
  \item The metallicity of the gas is lower than $10^{-4}Z_{\odot}$, 
\end{itemize}

where $n_{\rm Jeans}\equiv{\pi c_s^2}/\left({G\mu m_{\rm H}(8 R_{\rm cell})^2}\right)$ is the numerical Jeans number density, and $Z_{\odot}$ is solar metallicity. A density floor of $1\, {\rm cm^{-3}}$ is imposed to prevent spurious star formation in the low-density, low-temperature IGM. Such a Jeans-based density criterion follows the approach of THESAN-ZOOM \citep{kannan25}, but we adopt a more conservative prefactor of 0.5. This criterion is designed to be less sensitive to numerical resolution compared to the fixed density thresholds used in \citet{wise12}. In particular, such a criterion effectively incorporates the condition of local gravitational instability, since the density threshold is evaluated relative to the resolved gas properties rather than being a fixed global value. Simulation tests show that the results are robust to variations in this value (see Appendix \ref{appx:sfcrit}). The density threshold is resolution-dependent to ensure that star formation converges across different resolution levels.

The criterion for the ${\rm H_2}$ fraction is based on \citet{tegmark97}, who found that an ${\rm H_2}$ fraction of $\sim10^{-4}-10^{-3}$ is required for primordial gas to cool efficiently within a Hubble time. We adopt $5\times10^{-4}$ ,following \citet{wise12}'s value in their work. The metallicity criterion is set to $10^{-4}\, {\rm Z_\odot}$, motivated by early studies on the fragmentation of star-forming clouds \citep{bromm01, schineider02}. While later studies suggested that the critical metallicity for fragmentation could drop to $10^{-5}-10^{-6}\, {\rm Z_{\odot}}$ if dust cooling is considered \citep{omukai05, schneider06}, our current framework does not explicitly track dust evolution. Therefore, we choose a conservative limit of $10^{-4}\,{\rm Z_\odot}$. The potential impact of this metallicity criterion will be investigated in future work.

Once a gas cell satisfies these conditions, the star formation rate (SFR) is calculated as:
\begin{equation}\label{sfr}
{\rm SFR} = \varepsilon_\star\cdot m_{\rm Gas}/\tau_{\rm ff},
\end{equation}
where $\varepsilon_{\star}$ is the star formation efficiency (SFE) and $\tau_{\rm ff}\equiv\sqrt{3\pi /(32G\rho)}$ is the free-fall timescale of the gas. 

The local SFE within a gas cloud is physically regulated by feedback from protostellar outflows, radiation from massive stars, and supernovae. Observational and theoretical studies of metal-enriched star formation indicate that these processes typically limit the SFE to a few percent up to $\sim 10\%$ \citep[e.g.,][]{kennicutt2012, grudic22}. Theoretical works further suggest that Pop III star formation is similarly self-regulated by radiative feedback \citep{stacy16, jaura22} and potentially magnetic outflows \citep{machida06}. The resulting SFE depends on the interplay of these processes and the cloud properties. We performed comparison runs with $\varepsilon_\star=10\%$ and $\varepsilon_\star=1\%$. We found that reducing $\varepsilon_\star$ to $1\%$ results in a moderate decrease in the total Pop III stellar mass (to $\sim2/3$ of the fiducial value), indicating a sub-linear dependence on the local efficiency parameter. Meanwhile, the total mass of Pop II stars is also not significantly affected at $z=10$ (see Appendix \ref{appx:sfcrit}). Using cloud-size high-resolution simulations, \citet{jaura22} find that a cloud of $\sim 2700\,{\rm M_{\odot}}$ yields a total stellar mass of several tens to $\sim 100\,{\rm M_{\odot}}$, corresponding to an SFE of a few percent.  For simplicity, we adopt a fixed fiducial value of $\varepsilon_\star=10\%$ for both Pop III and Pop II star formation. 

Regarding the numerical implementation, we follow the stochastic approach of \citet{springel03}: a gas cell is converted into a star particle with a probability $p = 1 - \exp(-\varepsilon_\star dt / \tau_{\rm ff})$, where $dt$ is the timestep of the gas cell. The mass of the newly formed star particle corresponds to the mass of the original gas cell. In our fiducial high-resolution runs, the typical stellar particle mass is $\sim 100-200\,{\rm M_{\odot}}$. Given our adopted Pop III IMF with a mean mass of $\sim28\,{\rm M_{\odot}}$ (see Sec.~\ref{sec:imf}), each particle effectively represents a small multiple system of Pop III stars, whereas for Pop II stars, it represents a small stellar cluster. In cases where the star particle mass falls below $\sim30\,{\rm M_{\odot}}$, the particle should still be interpreted as a statistical sampling unit of the IMF, rather than an individual physical star. No minimum stellar particle mass is imposed in the model, and the associated radiative and supernova feedback are injected according to IMF-averaged expectations in a stochastic sense.




\subsubsection{The IMF of Pop III stars}\label{sec:imf}

The IMF of Pop III stars remains highly uncertain and is still a subject of active debate \citep{Klessen23}. While early theories suggested that Pop III stars formed in isolation as extremely massive objects \citep[e.g.,][]{hirano14}, more recent high-resolution simulations indicate that the primordial accretion disk is likely to fragment, leading to a broader mass distribution \citep{stacy16, jaura22}.

Since the primary aim of this paper is to introduce and evaluate our sub-grid framework for Pop III star formation and feedback, we adopt a representative and physically motivated IMF, rather than exploring the full parameter space of possible Pop III IMFs. Specifically, we use the fitting formula from \citet{jaura22}, which corresponds to a scenario where the star-forming region undergoes fragmentation. The fitting formula of the IMF is
\begin{equation}\label{eq:imf}
{\rm \Phi}(M) = \exp\left(-a(1/M)^b\right)M^{-c},
\end{equation}
where $\Phi(M)\equiv dN/dM$ represents the IMF, the fitting parameters are $a=2.1419,\ b=0.3920,\ c=1.1670$. The fitting parameters $a$, $b$, and $c$ are obtained by fitting to the IMF data presented in Figure 6 of \citet{Klessen23}, which is based on the simulation of \citet{jaura22}. The IMF is normalized such that the integral over the mass range yields a unit mass. The lower and upper mass limits are set to $1\,{\rm M_{\odot}}$ and $100\,{\rm M_{\odot}}$, respectively. The IMF-weighted Pop III stellar mass is $\sim28\,{\rm M_{\odot}}$. A comprehensive investigation into the impacts of different Pop III IMF slopes and mass limits will be presented in future work.

Regarding the numerical representation of stars, classic cosmological simulations typically assume that each star particle represents a Simple Stellar Population (SSP), as the mass resolution of these simulations is much higher than that of individual stars \citep{vogelsberger14, schaye15, pillepich18}. As computational power increases, the resolution of modern simulations can reach down to $\sim1\,{\rm M_{\odot}}$, making it possible to perform galaxy or cosmological simulations that resolve individual stars \citep{deng24, brauer2025}. Such star-by-star simulations provide a more physically grounded description of stellar feedback compared to the IMF-weighted average approach used in SSP models. However, achieving the requisite mass resolution remains computationally prohibitive for large parameter space studies or massive systems. In this work, our star particles (typical median mass $\sim 100-200{\rm M_{\odot}}$) represent small stellar clusters (or multiple stellar systems) sampled from the IMF. To bridge the gap between SSP treatments and single-star simulations, we implement a discrete feedback model. As described in Section \ref{sec:sfb}, SN energy is released through stochastic Poisson sampling rather than continuous injection, similar to the approach adopted in FIRE-2 \citep{hopkins2018fire2} and SMUGGLE \citep{marinacci2019}. This approach allows us to capture the stochastic nature of feedback while maintaining the computational efficiency of a subgrid framework suitable for lower resolutions.

\subsubsection{Pop II star formation}\label{sec:pop2sf}

Following the death of Pop III stars via core-collapse supernovae (CCSNe) or pair-instability supernovae (PISNe), metals are enriched into the surrounding environment. Subsequently, Pop II stars form in these metal-enriched regions. A gas cell is eligible to form a Pop II star particle if it meets the following criteria:

\begin{itemize}
  \item The gas is convergent $\nabla\cdot v_{\rm Gas} <0$;
  \item The gas number density $n_{\rm Gas}$ exceeds a threshold $n_{\rm th}$ of $\max(10\, {\rm cm^{-3}}, 0.5\, n_{\rm Jeans})$; 
  \item The metallicity of the gas exceeds the critical threshold of $10^{-4}Z_{\odot}$;
  \item The gas temperature is lower than $10^3\, {\rm K}$.
\end{itemize}
We adopt a higher density threshold $10\,{\rm cm^{-3}}$ compared to the Pop III case, reflecting the efficient cooling provided by metals, which allows the gas to collapse to higher densities before fragmentation. The calculation of the star formation rate (SFR) and the stochastic numerical implementation follow the same methodology as described for Pop III stars in Section \ref{sec:pop3sf}. 

In the simulation, each Pop II star particle represents a small stellar cluster following a Chabrier IMF \citep{chabrier03}. The lower and upper stellar mass limits are set to 0.1 ${\rm M_{\odot}}$ and 40 ${\rm M_{\odot}}$, respectively.

\subsection{Kinetic feedback from stars}\label{sec:sfb}

In this work, we incorporate kinetic feedback mechanisms, including supernova (SN) explosions from both massive Pop III and Pop II stars, as well as stellar winds from massive Pop II stars. Due to their pristine composition (i.e., lack of metals), Pop III stars are not expected to experience the strong classical stellar winds characteristic of massive stars, which are driven by radiation pressure on metal spectral lines \citep{castor75}. So we do not include stellar wind feedback for Pop III stars in our current framework.

\subsubsection{Evolutionary track of Pop III stars}

The evolutionary fate of Pop III stars is primarily determined by their initial stellar mass \citep[e.g.][]{heger02, Klessen23}. Based on current stellar evolution theories, the endpoints of Pop III stars can be categorized as follows:

\begin{itemize}
    \item Stars with masses in the range $10\,{\rm M_{\odot}} \le M_\star \le 25\,{\rm M_{\odot}}$ end their lives as CCSNe, typically releasing a kinetic energy of $E_{\rm SN} \sim 10^{51}\,{\rm erg}$.
    \item Stars in the range $25\,{\rm M_{\odot}} < M_\star \le 40\,{\rm M_{\odot}}$ generally experience significant fallback, resulting in faint supernovae with explosion energies far below $10^{51}\, {\rm erg}$, or direct collapse into black holes.
    \item Stars with masses between $40\,{\rm M_{\odot}}$ and $\sim 140\,{\rm M_{\odot}}$ are expected to collapse into BHs, either directly or following mass loss via pulsational pair-instability (PPISN).
    \item Stars with masses in the range $140\,{\rm M_\odot} \lesssim M_\star \lesssim 260\,{\rm M_\odot}$ undergo pair-instability supernovae (PISNe), which completely disrupt the star without leaving a remnant.
    \item Stars more massive than $260\,{\rm M_{\odot}}$ collapse directly into BHs as photodisintegration prevents the pair-instability explosion.
\end{itemize}

In this work, we adopt an upper mass limit of $100\,{\rm M_{\odot}}$ for the Pop III IMF, and therefore PISNe are not included. This choice is made to reduce the complexity of the feedback channels and to focus on the stability and convergence of the Pop III star formation model itself. We leave a more comprehensive treatment including PISNe for future work, with the variation of the IMF. We assign a canonical explosion energy of $E_{\rm SN} = 10^{51}\, {\rm erg}$ to stars in the mass range $10-25\,{\rm M_{\odot}}$. For stars outside this range (i.e. $> 25\,{\rm M_{\odot}}$), we assume they collapse directly into black holes without releasing feedback energy. Although recent observations of metal-poor stars suggest possible chemical imprints of PISNe \citep[e.g.][]{xing23}, we reserve the investigation of PISN feedback and yields for future work.

\subsubsection{SN feedback injection}

Instead of releasing supernova energy instantaneously from a newly formed star particle, we account for the finite lifetimes of stars with different masses. The SN energy release rate, $\dot{E}_{\rm SN}$, is calculated based on stellar evolution theory \citep{maraston05}:

\begin{equation}\label{eq:energy_rate} 
\dot{E}_{\rm SN} = m_* \Phi(M_{\rm TO}) \left| \frac{d M_{\rm TO}}{dt} \right| E_{\rm SN}, 
\end{equation}
where $m_*$ is the mass of the star particle, $\Phi(M)$ is the IMF normalized to unit stellar mas, and $M_{\rm TO}(t_{\rm age})$ is the main-sequence turn-off mass at the current age of the star particle $t_{\rm age}$. The turn-off masses for Pop III stars are interpolated from the tables provided by \citet{Klessen23}, which rely on the stellar evolution models of \citet{murphy21} and \citet{ekstrom12}. For Pop III CCSNe, we adopt a canonical explosion energy of $E_{\rm SN}=10^{51}\, {\rm erg}$.

Numerically, the SN energy release at each timestep $dt$ in the simulation can be calculated by $\dot{E}_{\rm SN} dt$. However, such an injection may make the energy injection lower than the energy released from one single SN event, especially when the mass resolution is high. To address this, we employ a stochastic injection method. At each timestep dt, the expected number of SN events, $\lambda=(\dot{E}_{\rm SN}\,dt)/E_{\rm SN}$, is calculated. We then determine the discrete number of SN events, $N_{\rm SN}$, by sampling from a Poisson distribution with mean $\lambda$. The total energy injected in that timestep is then $E_{\rm inj} = N_{\rm SN}E_{\rm SN}$. This ensures that energy is always injected in integer multiples of a single supernova event.

The coupling of feedback energy to the gas depends on whether the Sedov–Taylor phase of the supernova remnant (SNR) is resolved. We define two characteristic radii: the cooling radius $R_{\rm cool}$ \citep{li15}, where radiative losses become significant, and the fade-away radius $R_{\rm fade}$ \citep{li20}, where the SNR merges with the background medium. These are calculated as:

\begin{equation} \label{eq:coolingradius}
R_{\rm cool} = 23.7\, N_{\rm SN}^{0.29}\, n^{-0.42}\, (Z/{\rm Z_{\odot}}+0.1)^{-0.5}\,{\rm pc}, 
\end{equation}

\begin{equation} 
R_{\rm fade} = 48.8\, N_{\rm SN}^{1/3}\, \left(\frac{n}{0.02\,{\rm cm^{-3}}}\right)^{-1/3}\, \left(\frac{T}{10^7\rm K}\right)^{-1/3}\,{\rm pc}, 
\end{equation}

where $n$, $T$ and $Z$ denote the number density (in units of $\rm cm^{-3}$), temperature, and metallicity of the gas cell hosting the star particle. The metallicity-dependent factor $(Z/{\rm Z_{\odot}}+0.1)^{-0.5}$ in Eq. \ref{eq:coolingradius} accounts for reduction of radiative cooling at lower metallicities, which is derived based on \citet{draine2011}. We compare the cell radius, $R_{\rm cell}$ to the effective feedback radius $R_{\rm fb} = \min(R_{\rm cool}, R_{\rm fade})$. 

If $R_{\rm cell} < R_{\rm fb}$, the Sedov-Taylor phase is considered resolved. In this case, we inject the total feedback energy $E_{\rm inj}$ as thermal energy into the host gas cell.

If $R_{\rm cell} > R_{\rm fb}$, the Sedov-Taylor phase is unresolved, and the injection method depends on whether the cooling is efficient enough:

\begin{itemize}
    \item \textbf{Case 1 ($R_{\rm cool} < R_{\rm fade}$):} The SNR cools radiatively before fading, entering the momentum-conserving phase. We therefore inject the terminal momentum, $p_{\rm terminal}=\sqrt{2m_{\rm cool}E_{\rm inj}}$, distributed to the 32 nearest neighbors, where $m_{\rm cool}=4/3\pi\rho_{\rm cell} R_{\rm cool}^3$, and $\rho_{\rm cell}$ is the gas density of the host cell \citep{thornton98,kimm14,nunez17}. 
    \item \textbf{Case 2 ($R_{\rm cool} > R_{\rm fade}$):} The SNR merges with the background while still in the hot, adiabatic phase. In this regime, we inject $70\%$ of $E_{\rm inj}$ as thermal energy into the host cell and $30\%$ as kinetic energy into the 32 nearest neighbors, mimicking the energy partition of a Sedov-Taylor blast wave \citep{chevalier74, li15,nunez17}.
\end{itemize}

For Pop II stars, we similarly include CCSNe feedback for stars with masses $M_\star>8\,{\rm M_{\odot}}$, utilizing the same numerical implementation as described above but adopting the standard Chabrier IMF \citep{chabrier03}.

\subsubsection{Stellar winds from Pop II stars}

We model the stellar winds from massive Pop II stars following the prescriptions of \citet{vink2001}. Although the original model explicitly depends on metallicity, $\dot{M}\sim (Z/{\rm Z_{\odot}})^{0.85}$, in this work we adopt the mass-loss rates and velocities corresponding to solar metallicity as a simplified treatment since we find that the stellar wind is subdominant feedback channel. A more precise, metallicity-dependent treatment of stellar winds will be investigated in future studies. The adopted mass-loss rate, $\dot{M}_{\rm wind}$, is given by:

\begin{equation}
\log_{10}(\dot{M}_{\rm wind})=
    \left\{
\begin{aligned}
\begin{multlined}
        -6.688 + 2.21 \log(L_\star/10^5) \\
        - 1.339 \log(M_\star/30) \\
        - 1.601 \log(1.3/2.0) \\
        + 1.07 \log(T_{\rm eff}/(2\times10^4))
    \end{multlined},\ 
    & \quad \text{if } T_{\rm eff} < T_c \\
\begin{multlined}
        -6.697 + 2.194 \log(L_\star/10^5) \\
        - 1.313 \log(M_\star/30) \\
        - 1.226 \log(2.6/2.0) \\
        + 0.933 \log(T_{\rm eff}/(4\times10^4)) \\
        - 10.92 [\log(T_{\rm eff}/(4\times10^4))]^2
    \end{multlined},\ 
    & \quad \text{if } T_{\rm eff} > T_c
\end{aligned}
\right.
\end{equation}

where the wind velocity, $v_{\rm wind}$, is defined as:

\begin{equation}
{v}_{\rm wind}=
    \left\{
    \begin{aligned}
        1.3v_{\rm esc},\ & \text{if}\ \  T_{\rm eff} < T_c\\
        2.6v_{\rm esc},\ & \text{if}\ \  T_{\rm eff} > T_c
    \end{aligned}
    \right.
\end{equation}

Here, $L_{\star}$, $M_{\star}$, and $T_{\rm eff}$ denote the stellar luminosity, mass, and effective temperature in units of ${\rm L_{\odot}}$, ${\rm M_{\odot}}$ and Kelvin, respectively. $v_{\rm esc}\equiv\sqrt{2GM_\star/R_\star}$ is the escape velocity on the surface of stars, and $T_{\rm c}=25882\,{\rm K}$ represents the critical temperature of the bi-stability jump. 

For a single star particle representing a stellar population, the total wind momentum injection rate, $\dot{P}_{\rm sw}$, is calculated by integrating over the surviving massive stars:

\begin{equation}
    \dot{P}_{\rm sw} = \int^{M_{\rm TO}(t_{\rm age})}_{M_{\rm min}} \Phi(M) \left(\dot{M}_{\rm wind}(M)v_{\rm wind}(M)\right)dM,
\end{equation}
where $\Phi(M)$ is the IMF and $M_{\rm TO}(t_{\rm age})$ is the main-sequence turn-off mass at the current age of the star particle. In the simulation, this momentum flux is injected spherically into the 32 nearest gas neighbors. 

\subsubsection{Dipole correction}

The feedback energy or momentum from supernovae and stellar winds is distributed to surrounding gas cells according to their local volumes. However, a purely volume-weighted coupling cannot guarantee isotropy of the injected fluxes. Anisotropic injection may bias the deposited momentum toward certain directions, leading to unphysical bipolar outflows. \citet{hopkins18} demonstrated that such anisotropies can significantly distort galactic morphologies and even disrupt thin disks at high resolution. To alleviate this issue, they proposed a tensor renormalization scheme to enforce isotropy. 

In this work, we adopt a simpler dipole correction to ensure that the injected feedback momentum has no net preferred direction. For the weights assigned to neighboring gas cells $w_i$, the correction enforces
\begin{align}
\sum_i w_i &= 1, \\
\sum_i w_i \mathbf{e}_i &= \mathbf{0},
\end{align}
where $\mathbf{e}_i$ is the unit vector from the source star to the target gas cell. The initial (uncorrected) weights are defined as $w_i^{(0)} \equiv V_i/\sum_j V_j$, where $V_i$ is the gas-cell volume. We then apply a minimal correction,
\begin{equation}
w_i = w_i^{(0)}\left(1 - \mathbf{e}_i \cdot \boldsymbol{\lambda}\right),
\end{equation}
where the correction vector $\boldsymbol{\lambda}$ is obtained by solving the linear system
\begin{equation}
\mathbf{M}\cdot\boldsymbol{\lambda} = \mathbf{d},
\end{equation}
with $\mathbf{M} \equiv \sum_i w_i^{(0)}\,\mathbf{e}_i\mathbf{e}_i^{\mathrm{T}}$ and $\mathbf{d} \equiv \sum_i w_i^{(0)}\,\mathbf{e}_i$. 
This system always has a unique solution as long as the local cell distribution spans three dimensions. 

Although this correction guarantees momentum isotropy, it does not strictly ensure $w_i>0$. In rare cases where negative weights appear (typically for highly anisotropic neighbor configurations), we clip the negative values to zero and renormalize the weights. This slightly relaxes the $\sum_i w_i \mathbf{e}_i = 0$ constraint, but the residual dipole is negligible in practice. Compared to the full tensor renormalization of \citep{hopkins18}, our approach is simpler to implement and sufficiently accurate for supernova feedback modeling, leaving only a minor quadrupole residual. 

\subsection{Mass return and metal yields}\label{sec:metal}

Stars lose a significant fraction of their initial mass over their lifetime, through main-sequence stellar winds, post-main-sequence mass loss, and explosive ejection at the end of their evolution. However, the certain fraction in different period depends on the metallicity and stellar mass. To maintain feasibility, we adopt a simplified model where the total mass loss is returned to the interstellar medium (ISM) in a single instantaneous event at the end of the life of the stars. This approach, while simplifying the precise timing, captures the crucial feedback of the total mass and metals returned. We argue that any uncertainties introduced by this timing approximation are likely subdominant to the larger, more fundamental uncertainties inherent in modeling the early universe, such as the unknown IMF of Pop III stars. 

Adopting treatment from some previous works \citep{novak11, yuan18}, the mass return rate from a stellar population is calculated as:

\begin{equation}\label{massloss}
\dot{M}_{\rm return} = m_*{\Phi}(M_{\rm TO})\left|\frac{d {M}_{\rm TO}}{dt} (t_{\rm age})\right|\Delta M,
\end{equation}
where $\Delta M \equiv M_{\rm ZAMS}-M_{\rm rem}$ is the mass difference between the mass at ZAMS and the remnant of the star with mass $M_{\rm TO}(t_{\rm age})$. 

The metal yield rate is calculated using a similar formulation \citep{zhu23}:
\begin{equation}\label{metalloss}
\dot{M}_{\rm Z} = m_*{\Phi}(M_{\rm TO})\left|\frac{d {M}_{\rm TO}}{dt} (t_{\rm age})\right|m_{\rm Z,\, yields},
\end{equation}
where $m_{\rm Z,\, yields}$ is the metal return during the evolution of the star. In this work, $M_{\rm rem}$ and $m_{\rm Z,\, yields}$ is obtained from the Yields Table in \citet{nomoto13}. The injection of the mass return and metal yields follows the method of SN injection described in Section \ref{sec:sfb}.

\subsection{Radiation feedback}\label{sec:rfb}

Instead of implementing full RT, which is computationally expensive, we employ a hybrid approach combining a short-range approximation with a long-range optical-thin tree-based RT method. The long-range optical-thin tree-based RT method is based on  \citet{zhu24}, which is based on \citet{kannan14}. This approach, while sacrificing some accuracy compared to full RT, is computationally efficient and shares similarities with the Locally Extincted Background Radiation in Optically Thin Networks (LEBRON) method \citep{hopkins20}.

We divide the stellar radiation into two bands: the Lyman-Werner (LW) band (11.2--13.6 eV) and the ionizing band (13.6--100 eV), to capture the effects of radiation on ${\rm H_{2}}$ dissociation and photoionization of the primordial gas. 

Since the IGM is largely transparent to LW radiation in the early Universe, we adopt an optically thin approximation for this band, following the methods described in \citet{zhu24}. While the birth cloud surrounding a newly formed star can initially provide strong ${\rm H_2}$ self-shielding against LW radiation, the star's own LW photons photodissociate the surrounding ${\rm H_2}$ on a timescale of order $\sim1–10 \,{\rm kyr}$, much shorter than the stellar radiation lifetime. Birth cloud shielding of LW is therefore a brief transient that does not significantly affect the long-range LW background. The radiation flux from each star particle is calculated as $F=L/4\pi r^2$. To accelerate the computation, radiative information is propagated via the gravitational tree. For the photodissociation of ${\rm H_2}$ by LW radiation, we implement the self-shielding factor $f_{\rm sh}$ from \citet{wg11}:
\begin{equation}
    f_{\rm sh} = \frac{0.965}{(1+x/b_5)^{1.1}}+\frac{0.035}{(1+x)^{0.5}}\exp\left[-8.5\times10^{-4}(1+x)^{0.5}\right],
\end{equation}
 where $x=N_{\rm H_2}/5\times10^{14}\, {\rm cm^{-3}}$, and $b_5=b/10^5 {\rm cm\,s^{-1}}$, $N_{\rm H_2}$ is the column density of ${\rm H_2}$, $b$ is the Doppler broadening parameter. The column density is calculated by the local approximation $n_{\rm H_2}\cdot L_{\rm sob}$, where $L_{\rm sob}$ is the Sobolev-like length defined as $\rho/|\nabla \rho|$ \citep{gnedin2009}. 

In contrast to LW radiation, the early Universe is not optically thin to ionizing radiation. Ionizing photons form a Strömgren region around sources, beyond which the radiation field decays rapidly. We therefore assume that ionizing radiation is confined to this region. To determine the size of the ionized region, we perform an iterative neighbour search starting from the radius of the host gas cell of the sources, doubling the search radius in each iteration until the recombination rate is larger than the source luminosity:

\begin{equation}
  \sum_i {V_i}n_{\rm H}^2\alpha > Q,
\end{equation}
where $\alpha\equiv8.4\times10^{-11}T^{-0.5}(T/10^3)^{-0.2}(1+(T/10^6)^{0.7})^{-1}$ is the recombination rate of hydrogen \citep{katz96}, $n_{\rm H}$ is the hydrogen number density, $Q$ is total ionizing photon release rate from the stars, and . While computationally efficient, this method assumes spherical symmetry and ignores the anisotropy of the surrounding gas distribution. We note that this corresponds to a Case A recombination rate. Since the iterative search doubles the radius at each step, the $\sim17\%$ difference relative to Case B is within a single search step and does not affect the resulting ionized region boundary in practice.

Once the ionizing region is defined, we calculate the 1D column density $N_{\rm H}(r)$ along the line of sight from the star. The local radiation flux is then given by $F=L/(4\pi r^2)\exp(-N_{\rm H} \sigma)$, where $\sigma$ is the SED-weighted ionization cross-section. Additionally, we account for the self-shielding of dense gas clumps within the ionized region. The shielded flux for cold clumps is calculated as $F_s=F\exp(-n_{\rm H} L_{\rm Jeans} \sigma)$, where $n_{\rm H}$ is local hydrogen number density, $L_{\rm Jeans}$ is the local Jeans length. More accurate models considering anisotropy will be explored in future work.

The Spectral Energy Distribution (SED) of Pop III stars is modeled as a blackbody spectrum. The effective temperatures are derived from the stellar evolution calculations of \citet{Klessen23}, which is based on \citet{murphy21}. We find that the difference between this blackbody approximation and the detailed photon rates is less than $10\%$. For each star particle, we integrate the SED over the adopted IMF.

For Pop II stars, the SED follows the treatment of young stellar populations in \citet{zhu24}, but we exclude the X-ray component (radiation from SN remnants and high-mass X-ray binaries). The SED is modeled using the Binary Population and Spectral Synthesis \citep[BPASS v2.3, ][]{eldridge17} models with solar metallicity at an age of 10 Myr, assumed to be unchanged with stellar age. We note that this approximation underestimates the time-integrated ionizing photon budget by a factor of $\sim3$ compared to a full temporal integration of the BPASS v2.3 model. Since our approximate 1D RT calculation already introduces significant uncertainties, this factor is likely subdominant within the current framework. In \citet{zhu24}, an escape fraction is set to mimic the shielding effect of star forming region and focus on the radiation on galaxy and CGM properties. In this work, we perform an approximate 1D radiative transfer within the ionized region, computing the attenuation from the neutral hydrogen column density $N_{\rm HI}(r)$ along the line of sight. The effective escape of ionizing photons is therefore determined self-consistently by the calculation, rather than prescribed as a subgrid parameter.

To maintain computational feasibility, we adopt a simplified temporal model where the SED of a stellar population is assumed to be constant over its main feedback phase. The radiation from Pop III populations is assumed to last for the first $10^{7.3}\, {\rm yr}$, corresponding to the lifetimes of 10 ${\rm M_{\odot}}$ stars. The radiation from Pop II populations is assumed to last for the first $10^{7.5}\,{\rm yr}$, corresponding to the lifetime of an $8\,{\rm M_{\odot}}$ star at solar metallicity. Although this neglects detailed temporal evolution, it captures the time-averaged radiative feedback, which is the dominant effect for the cosmological scales and long-term evolution studied here.
\begin{table*}
\centering
\caption{Summary of the simulations we performed, presented and analyzed in this paper. $L$ is the box size. $\epsilon_{\rm DM}$ and $\epsilon_{\rm Gas}$ is the gravitational softening length of DM and gas. $m_{\rm DM}$ is the mass resolution of the mass.}
\begin{tabular}{cccccc}
\hline
Simulation & $L\ [c{\rm Mpc}/h]$ & Total number of particles &  $\epsilon_{\rm DM}\, [c{\rm pc}]$ &  $\epsilon_{\rm Gas}\, [c{\rm pc}]$ & $m_{\rm DM}\ [{\rm M_{\odot}}]$ \\
\hline
IC0 & 1 &  $2\times 256^3$ & 100 & 17 &  $6.38\times10^3$\\
IC1 & 1 &  $2\times 256^3$ & 100 & 17 &  $6.38\times10^3$\\
IC2 & 1 &  $2\times 256^3$ & 100 & 17 &  $6.38\times10^3$\\
HRes & 1 &  $2\times 448^3$ & 100 & 17 & $1.88\times10^3$\\
LRes & 1 &  $2\times 128^3$ & 100 & 17 & $5.09\times 10^4$ \\
\hline
\end{tabular}
\label{tab:sim}
\end{table*}

\subsection{Turbulent metal mixing}\label{sec:mixing}

We include a subgrid model for the unresolved turbulent transport of metals in the interstellar and circumgalactic medium, following a Smagorinsky-type \citep{smagorinsky63} large eddy simulation (LES) approach \citep[e.g.][]{shen10,escala18}. In this framework, the evolution of the metal mass fraction $Z$ is governed by an advection–diffusion equation,
\begin{equation}
\frac{\partial \rho Z}{\partial t}+\nabla\cdot(\rho Z {\bf v}-\rho D \nabla Z) = 0,
\end{equation}
where $\rho$ and $v$ denote the gas density and velocity, and $D$ is the effective turbulent diffusion coefficient. We adopt a Smagorinsky-like prescription for the diffusion coefficient, 
\begin{equation}
D=C_{\rm diff} (2R_{\rm cell})^2 |S|
\end{equation}
where $R_{\rm cell}$ is the radius of the local Voronoi cell, $|S|$ is the norm of the traceless shear tensor of the velocity field, and $C_{\rm diff}$ is a dimensionless coefficient. The shear tensor is defined as
\begin{equation}
    S_{ij}=\frac{1}{2}\left(\frac{\partial v_i}{ \partial x_j}+\frac{\partial v_j}{\partial x_i}\right)-\frac{1}{3}(\nabla\cdot v)\delta_{ij}
\end{equation}
and its norm is computed as
\begin{equation}
    |S|=\sqrt{2S_{ij}S_{ij}}.
\end{equation}
Numerically, the diffusion equation is solved in a conservative form using an SPH-like pairwise discretization between neighbouring gas cells. We adopt 
$C_{\rm diff}$=0.15 as our fiducial value. We perform comparison simulations with and without this turbulent diffusion module, and find that the inclusion of turbulent metal mixing increases the volume filling fraction of metal-enriched gas, though the effect is comparable to the stochastic run-to-run variation (see Section III C and Appendix \ref{appx:random}). A systematic exploration of the dependence on the diffusion strength will be presented in future work.

\subsection{Chemistry and Cooling}\label{sec:clchem}

In the early Universe, the assumption of chemical equilibrium for primordial gas often breaks down as the relevant reaction timescales can be comparable to or longer than the dynamical timescales. Furthermore, molecular hydrogen $H_2$ acts as the primary coolant facilitating the formation of the first stars. We therefore employ a non-equilibrium chemical network for nine species: i.e., $\rm H^0$, $\rm H^+$, $\rm He^0$, $\rm He^+$, $\rm He^{++}$, $\rm H_2$, $\rm H_2^+$, $\rm H^-$, $\rm e^-$. The network includes 26 collisional and photo-ionization processes, with reaction rates adopted from \citet{katz96, abel97, glover08}. The chemical network includes the formation and destruction of atomic hydrogen, atomic helium and molecular hydrogen. We list these chemical reactions in Appendix \ref{appx:reaction}. We do not include deuterium chemistry, which becomes the dominant cooling channel at temperature $T<200\,{\rm K}$ and is essential for resolving the formation of individual Pop III stars. However, since the resolution of our current simulations is insufficient to resolve the collapse of individual protostellar cores, and this work focuses on larger-scale feedback effects, we neglect deuterium which is only important for direct Pop III star formation simulation.

Following previous treatment in cosmological simulations \citep{vogelsberger14, pillepich18, kannan25} $\Lambda_{\rm prim}$, metal cooling/heating $\Lambda_{\rm Z}$, and the Compton cooling due to the cosmic microwave background (CMB) $\Lambda_{\rm C}$:
\begin{equation}
    \Lambda = \Lambda_{\rm prim} + \Lambda_{\rm Z} + \Lambda_{\rm C}.
\end{equation}

The cooling rates for atomic primordial processes are taken from Table 1 of \citet{katz96}, which includes collisional excitation, collisional ionization, recombination, dielectric recombination, and free-free emission. For molecular cooling, we adopt the rates from Table 8 of \citep{glover08}, which assumes a fixed 3:1 ortho-to-para ratio for ${\rm H_2}$. This includes collisional cooling via the collision between $\rm H_2$ and $\rm H^0$, $\rm H_2$, $\rm He^0$, $\rm H^+$, and $\rm e^-$. We impose a temperature floor of $100\,{\rm K}$ for the cooling rates, as the neglect of deuterium chemistry renders the cooling function inaccurate below this threshold.

The heating rate due to photoionization and photodissociation is calculated as:
\begin{equation}
    \varepsilon_i=\int_{\nu_{\rm T}}^{\infty}\frac{4\pi J_{\nu}}{h\nu}\sigma_{\nu}(h\nu-h\nu_{\rm T}) d\nu,
\end{equation}
 where $i$ represent different processes, and $J_{\nu}$ is the intensity of the radiation field obtained by the method mentioned in Section \ref{sec:rfb}, $\nu_{\rm T}$ is the threshold frequency for the relevant process. 

We compute metal cooling/heating rates using {\sc Cloudy} \citep{ferland17} assuming chemical equilibrium. We generate a cooling/heating table, stored as HDF5 files. The table stores the metal cooling/heating rate $\Lambda_{\rm Z}=\Lambda_{\rm Z}(n_{\rm H}, T,Z, J_{\rm Pop\, III}, J_{\rm Pop\, II})$ that affected by stellar radiation, where $J_{\rm Pop\, III}$ and $J_{\rm Pop\, II}$ represent the intensity of stellar radiation from Pop III stellar population and Pop II stellar population.

The Compton cooling/heating rate due to CMB follows the previous works \citep{katz96}, which can be expressed as 
\begin{equation}
    \Lambda_{\rm C}=5.41\times10^{-36}n_eT(1+z)^4\, {\rm erg\,s^{-1}\,cm^{-3}}.
\end{equation}

\begin{figure*}
 \subfigure{\includegraphics[width=0.85\linewidth]{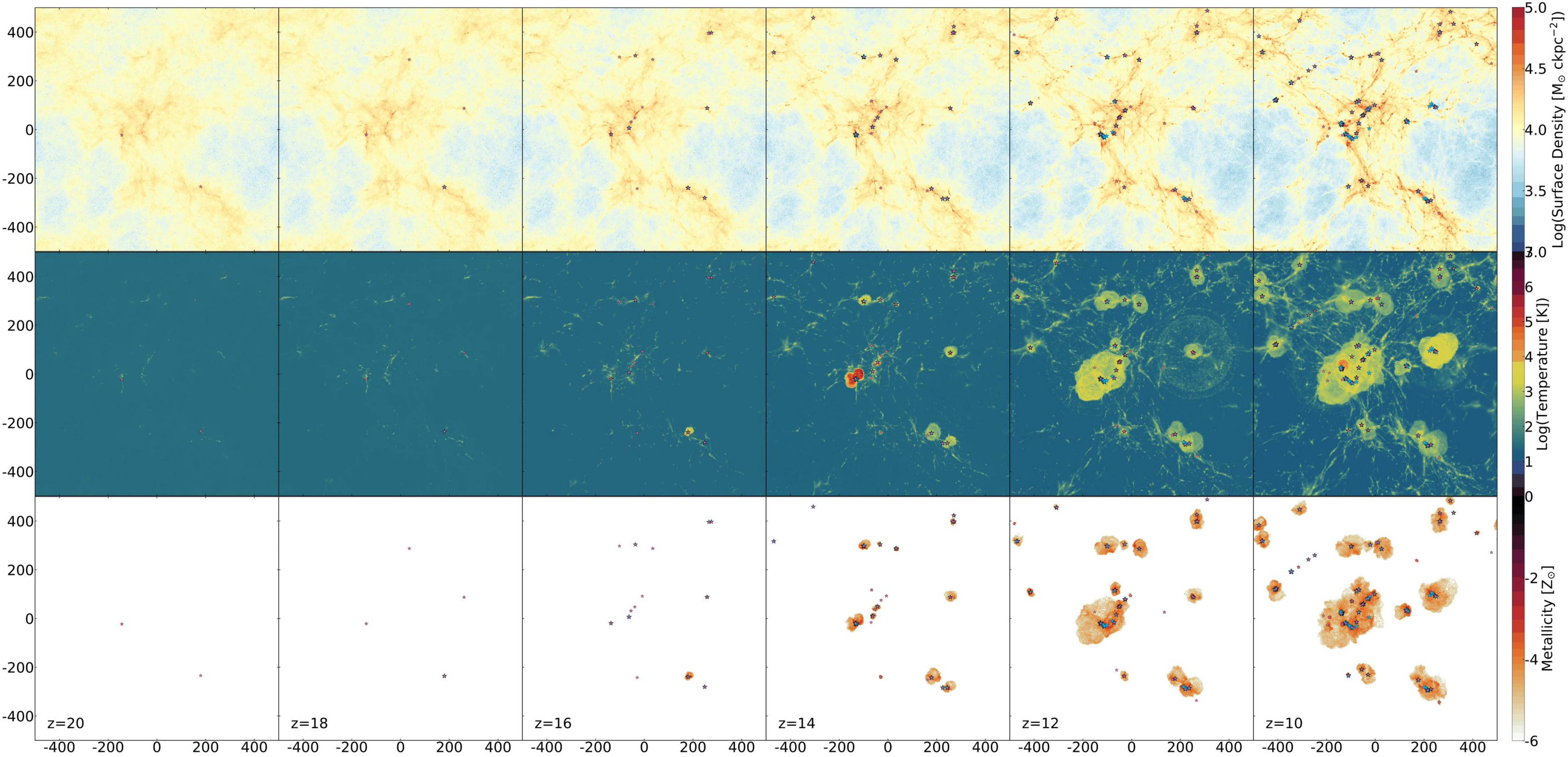}}\\
 \subfigure{\includegraphics[width=0.85\linewidth]{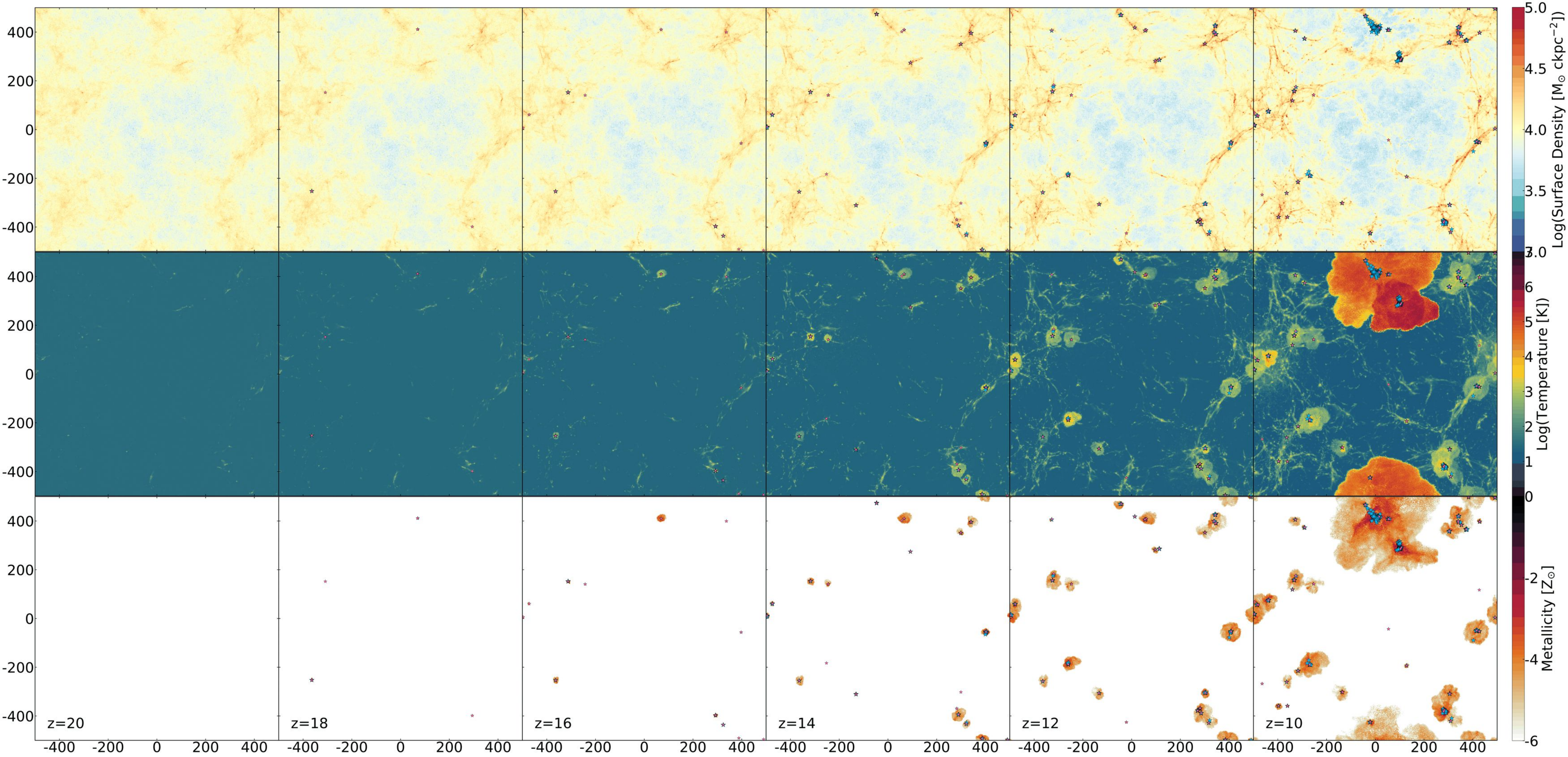}}\\
\subfigure{\includegraphics[width=0.85\linewidth]{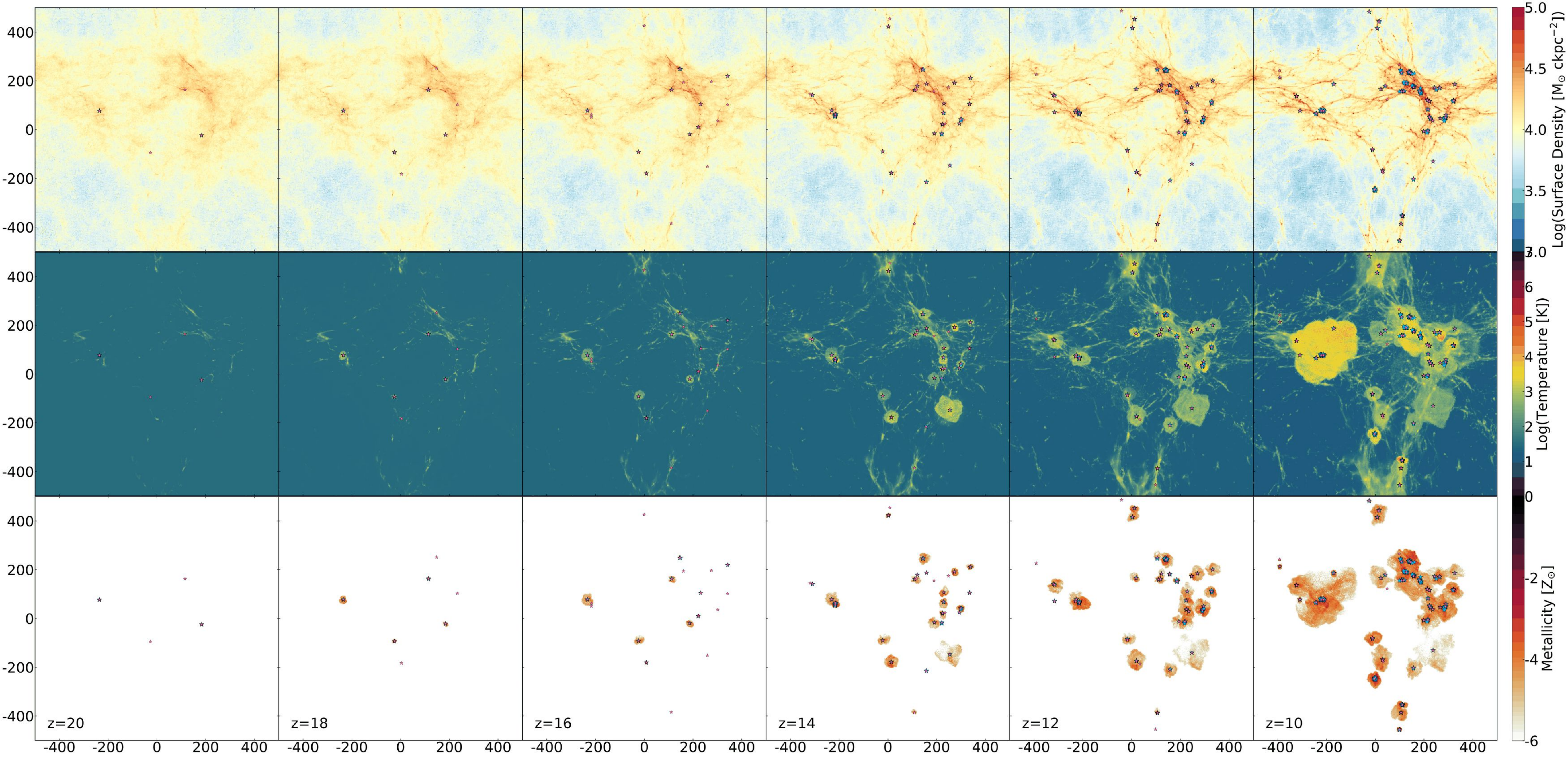}}\\
 \caption{The gas surface density, mass-weighted temperature, and metallicity distribution in the simulations with different initial conditions from $z=20$ to $z=10$. From the top to the bottom is IC0 (upper), IC1 (middle), and IC2 (bottom). The lime stars represent Pop III stars while the blue stars represent Pop II stars.} The first Pop III star forms at $z=20-18$ in these three simulations.
 \label{fig:std_dis}
\end{figure*}

\begin{figure}
 \subfigure{\includegraphics[width=1\linewidth]{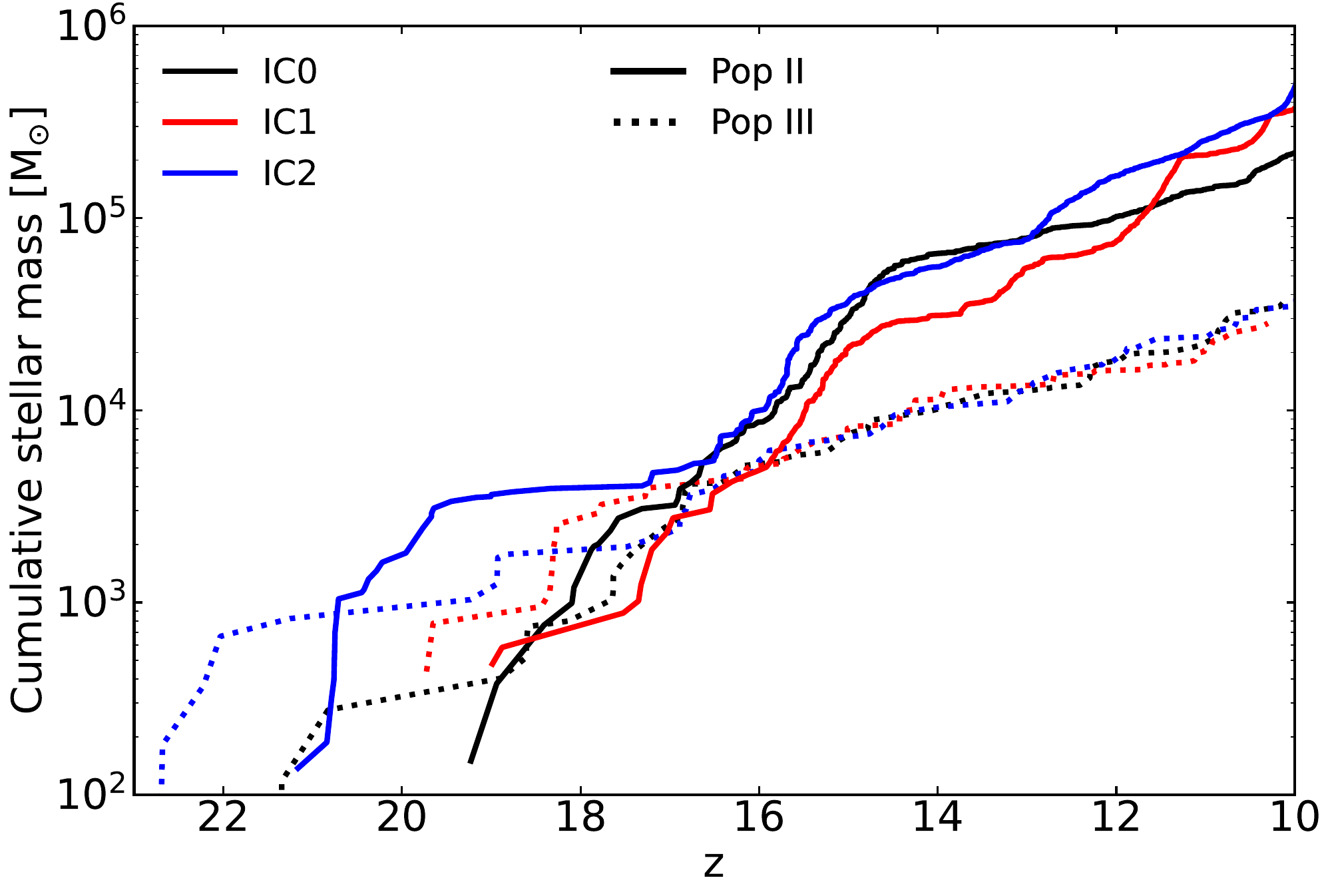}}\\
 \subfigure{\includegraphics[width=1\linewidth]{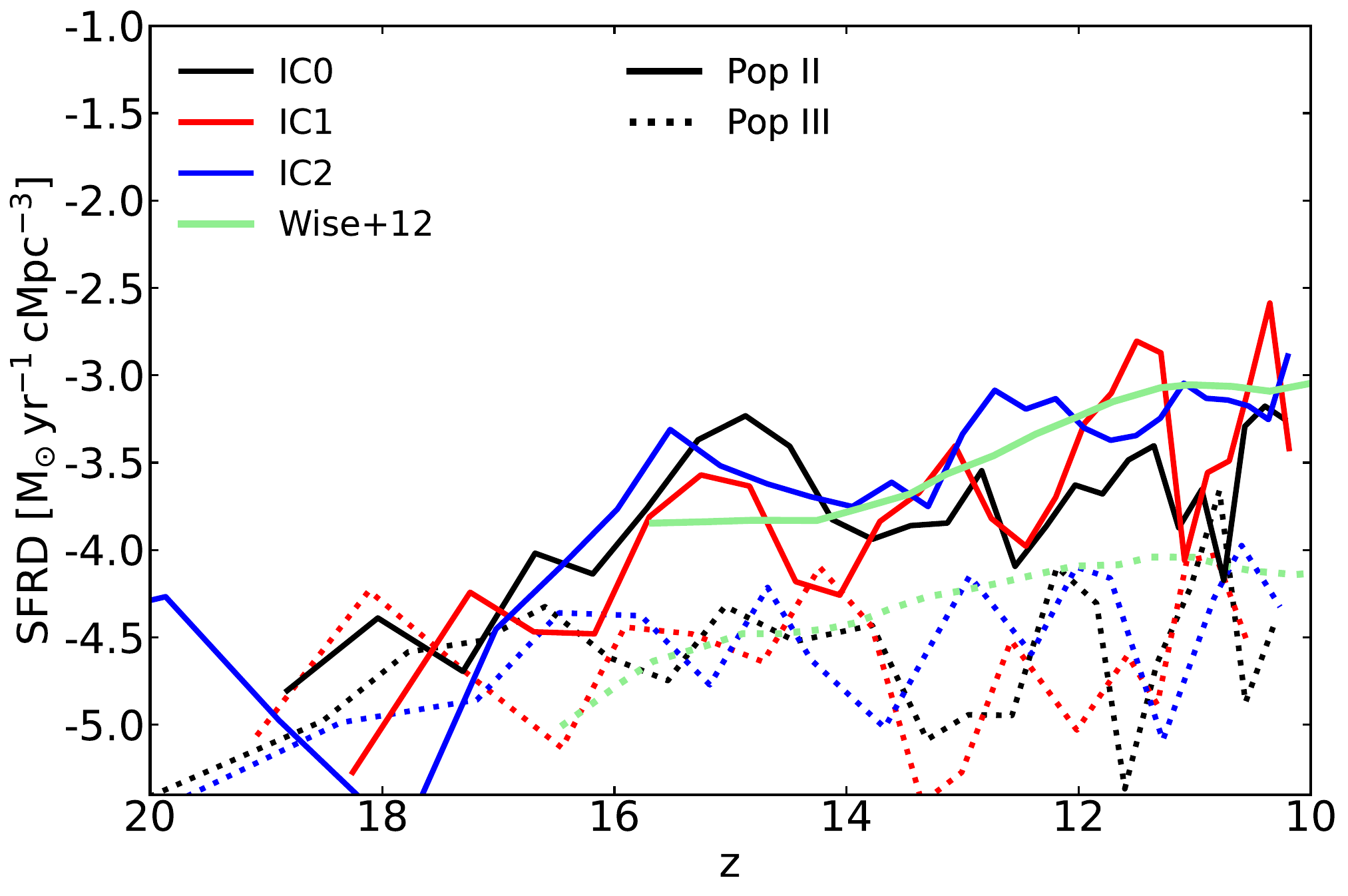}}
 \caption{{\it Upper:} The cumulative Pop III and Pop II stars in three simulations. The solid lines represent the Pop II star while the dashed lines represent the Pop III star. {\it Bottom:} The Pop III and Pop II star formation rate density (SFRD) from $z=20$ to $z=10$ in three simulations. 
 The green solid and dashed lines represent the simulations results about Pop II and Pop III SFRD in \citet{wise12}.}
 \label{fig:std_sfr}
\end{figure}

\subsection{Adaptive refinement strategy}\label{sec:refinement}

We adopt a Jeans-length based refinement criterion to ensure that the local Jeans scale is always resolved by at least four resolution elements (i.e. $N_{\rm J}\ge4$), preventing artificial fragmentation in dense regions \citep{truelove97,greif11,kannan25}. To avoid excessive refinement in very dense and cold regions, which would otherwise drive the cell size to extremely small values and significantly increase the computational cost, we set a temperature floor $T$ whenever the cell diameter falls below 4 pc. This floor does not affect the thermodynamics of already resolved gas, but simply prevents unlimited mesh refinement at high densities.

For de-refinement, we prevent merging of cells across strong gradients in density, temperature, velocity, or pressure. The tolerance factor is set to $f_{\rm gentle}=1.2$, meaning that a cell is excluded from de-refinement if its local physical quantities differ from the neighboring minimum or maximum by more than $20\%$. We further disable this gentle de-refinement criterion in two specific regimes:
\begin{itemize}
    \item {\bf Cold gas $T<3\times10^4\,{\rm K}$ and $M_{\rm gas} <100\,{\rm M_{\odot}}$:} We disable the gentle de-refinement constraint. This allows turbulent cold gas clouds to merge more aggressively. Since we do not aim to resolve the internal collapse of individual cold clumps below the Jeans floor, this reduces computational cost while maintaining consistency with our subgrid star-formation model (where the median gas mass is $\sim100-200\,{\rm M_{\odot}}$;
    \item {\bf Hot gas $T>3\times10^4\,{\rm K}$ and $M_{\rm gas} <10\,{\rm M_{\odot}}$:} We strictly prevent de-refinement for these gas cells. This ensures that the hot, supernova-driven outflow phase, which typically consists of low-mass, high-temperature cells, retains high mass resolution. This is crucial for accurately capturing the development and propagation of galactic outflows.
\end{itemize}

We tested the de-refinement criterion for these two specific regimes and obtain a convergent result with maximum computational efficiency.

\subsection{Simulation setup}\label{sec:simset}

In this work, we perform  a suite of cosmological hydrodynamical simulations in a periodic box with a comoving side length of $1\,{\rm cMpc}/h$, evolving from $z=127$ to $z=10$. 

Initial conditions are generated using the \textsc{N-GenIC} code \citep{springel05, angulo12}. The adopted cosmology follows the \textit{Planck} intermediate results \citep{planck16}, consistent with TNG: $\Omega_\Lambda=0.6911$, $\Omega_m=0.3089$, $h=0.6774$, $\Omega_b=0.0486$, $n_s=0.9667$, and $\sigma_8=0.8159$. The initial power spectrum is computed using the \citet{eisenstein99} transfer function. The default resolution employs $256^3$ dark matter particles and an equal number of gas cells in a $1\,{\rm Mpc}/h$ periodic box, corresponding to mass resolutions of $\sim1075\,{\rm M_\odot}$ for gas and $\sim6000\,{\rm M_\odot}$ for dark matter. However, since we include the gas cell refinement, the highest gas cell resolution will change during the simulation and can reach $\sim 10\, {\rm M_{\odot}}$. The gravitational softening length is 100 comoving pc for dark matter, while for gas cells and star particles, it is adaptive with a minimum of 17 comoving pc. 

In addition to the fiducial run, we perform two simulations with different random realizations of the initial conditions, as well as two resolution tests with particle masses of $3$ and $1/8$ times the fiducial runs. The main simulation parameters are summarized in Table~\ref{tab:sim}.

\section{Results}\label{sec:dic}

In this section, we present the global results from three simulations performed with different ICs but identical physical parameters and resolution. The time evolution of spatial distributions of gas density, temperature, and metallicity is presented in Section \ref{sec:rsd}. The global star formation rate density (SFRD), together with a comparison to \citet{wise12}, is shown in Section \ref{sec:rsf}. Finally, in Section \ref{sec:rgp}, we analyse the gas phase properties and the metal enrichment of the IGM driven by SN feedback.

\begin{figure*}
\includegraphics[width=\linewidth]{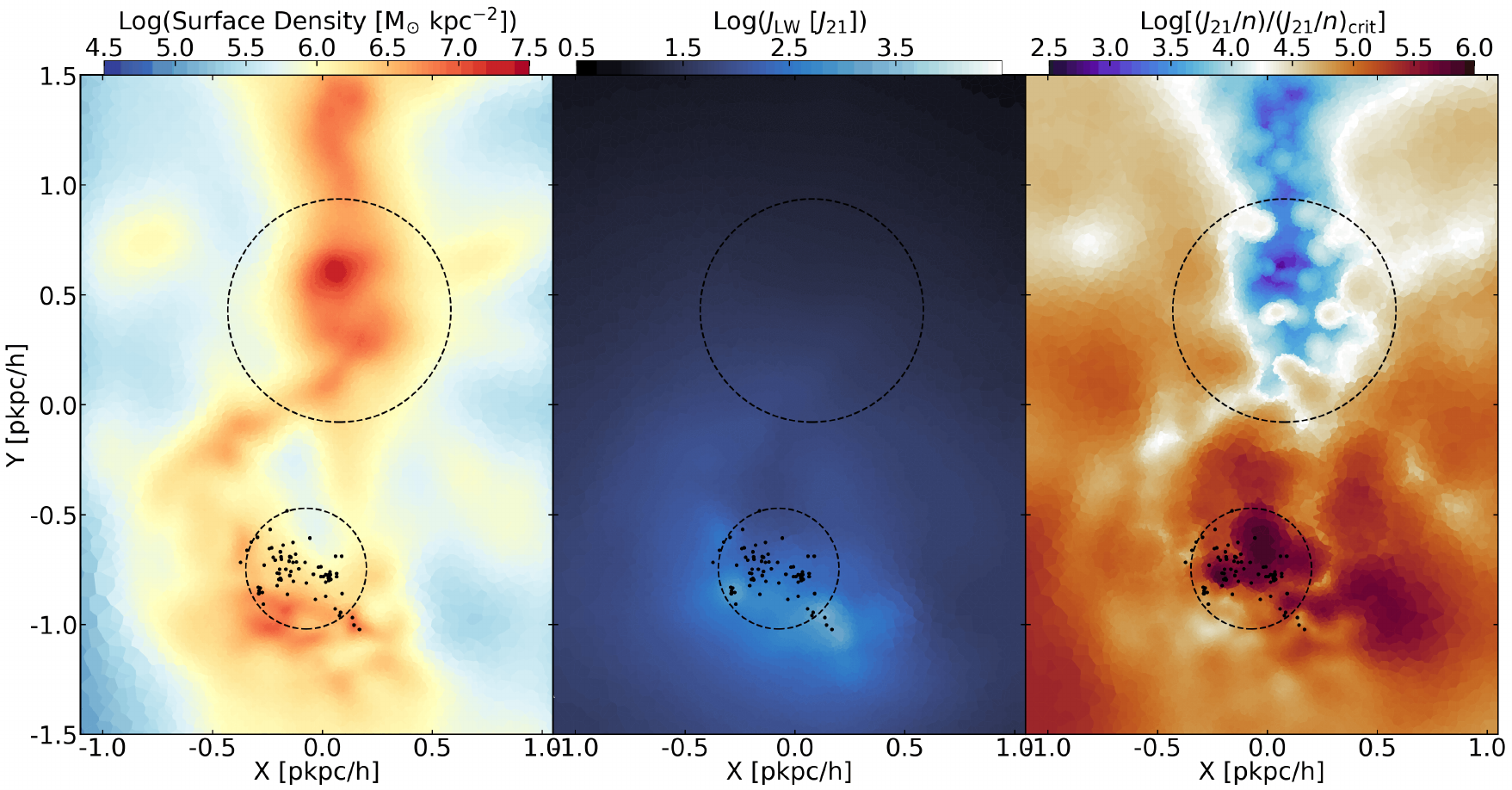}
 \caption{The gas surface density, LW radiation intensity ($J_{21}$), and the ratio of the LW intensity to the critical threshold for ${\rm H_2}$ cooling suppression for the two most massive subhalos in the IC0 simulation at $z=10$. The black dashed circles represent the radius containing half of the total mass, and black points indicate star particles. The third panel displays $\log\left( (J_{21}/n)/(J_{21}/n)_{\rm crit} \right)$; regions with values $>0$ indicate where the LW background is strong enough relative to the density to dissociate ${\rm H_2}$ and suppress cooling. This visualization explicitly confirms that the star-forming neighbor (bottom) generates a strong LW flux that irradiates the massive quiescent halo (top), preventing gas cooling and star formation in its center.}
 \label{fig:std_interaction}
\end{figure*}


\subsection{Time evolution of gas structure and metal enrichment}\label{sec:rsd}

Figure \ref{fig:std_dis} displays the time evolution of gas surface density, mass-weighted temperature, and metallicity in the three simulations (IC0, IC1, and IC2) from $z=20$ to $z=10$. The sequence of snapshots illustrates the formation of large-scale structures and the subsequent star formation history. 

In the IC0 and IC2 runs, dense structures form early, with the first Pop III stars already present at $z=20$. In the IC0 simulation, only two Pop III stars have formed by this epoch. However, due to the stochastic sampling of SN feedback described in Section \ref{sec:sfb}, these specific stars fall into mass ranges associated with direct collapse into black holes (or faint supernovae), resulting in no metal enrichment of the surrounding gas. In contrast, in the IC2 run, the Pop III stars fall into mass ranges associated with core-collapse supernovae, leading to metal enrichment of their surrounding gas. Consequently, Pop II stars begin to form from the metal-enriched gas. In the IC1 simulation, structure formation occurs somewhat later, with the first star appearing between $z=20$ and $z=18$. 

As dense structures continue to evolve, star formation proceeds continuously across the simulation volume. Supernova explosions drive hot bubbles that expand and transport metals to larger scales. These metal-enriched regions grow and gradually connect to each other. By $z=10$, the densest regions in all three simulations host a significant number of stars, driving large-scale, hot superbubbles. We discuss the detailed gas properties and metal enrichment processes in Section \ref{sec:rgp}.

In the IC0 simulation, the imprint of ionizing radiation feedback is clearly visible in the $z=12$ and $z=10$ snapshots. At $(x,\,y)\sim(250,\,100)\, c{\rm kpc}/h$, a slightly heated, spherical bubble with a radius of $\sim100\,c{\rm kpc}/h$ appears. This feature arises because the dense gas surrounding the source has been cleared, allowing ionizing photons to leak out from the small-scale region. The strictly spherical shape of this ionized bubble is an artifact of the one-dimensional approximation used for the column density calculation in our current radiation transport scheme; future work will incorporate more accurate angular-dependent column density calculations to capture anisotropic expansion.

\begin{figure*}
 \includegraphics[width=1\linewidth]{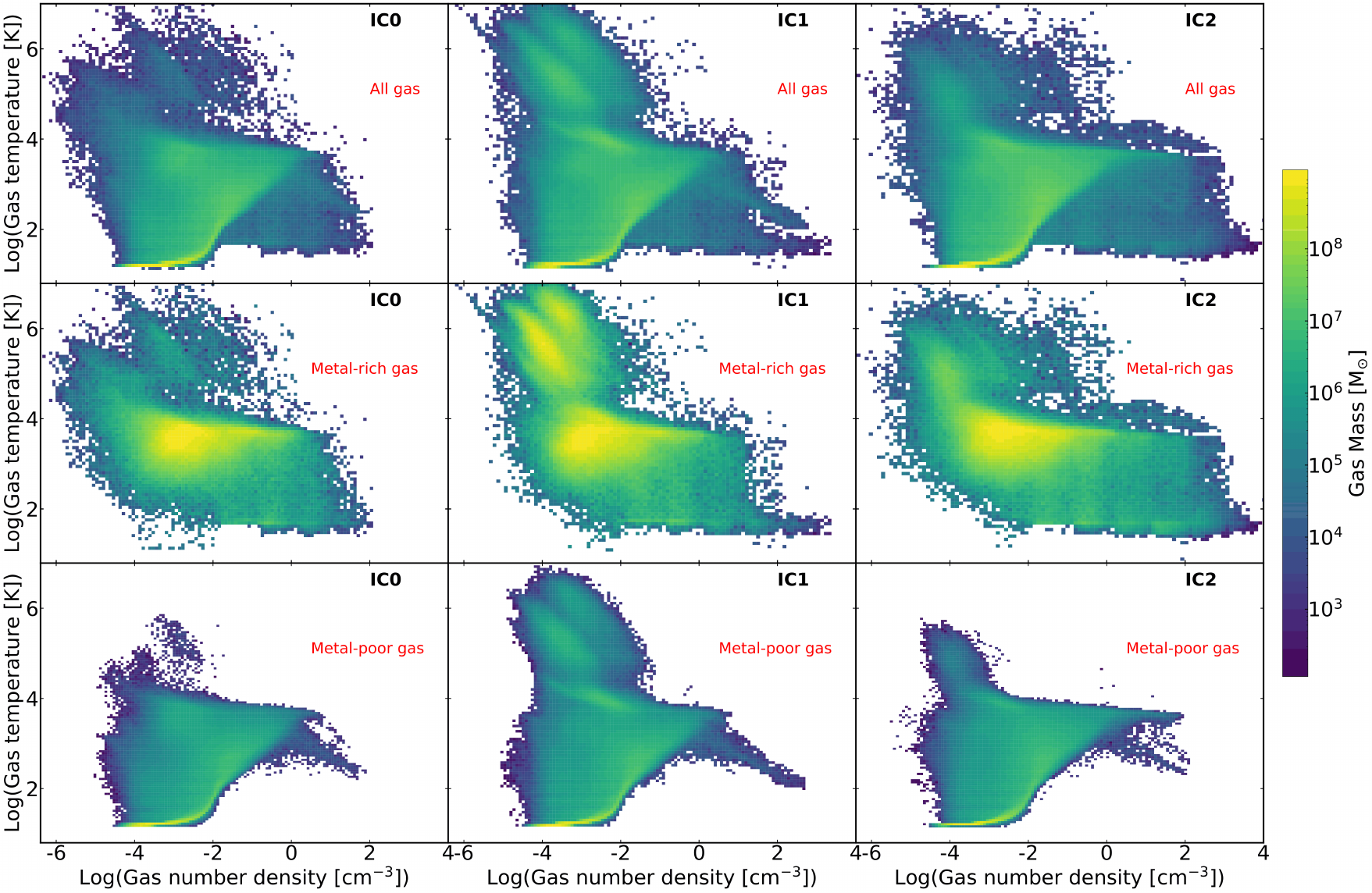}
 \caption{The gas phase diagram of number density and temperature for all gas ({\it upper}), metal-rich gas({\it middle}, defined by $Z/Z_{\odot}\ge10^{-4}$), and metal-poor gas ({\it bottom}, defined by $Z/Z_{\odot}<10^{-4}$)} at z=10. Each row represents the results from the same simulation.
 \label{fig:std_gas_phase}
\end{figure*}

\begin{figure}
\includegraphics[width=1\linewidth]{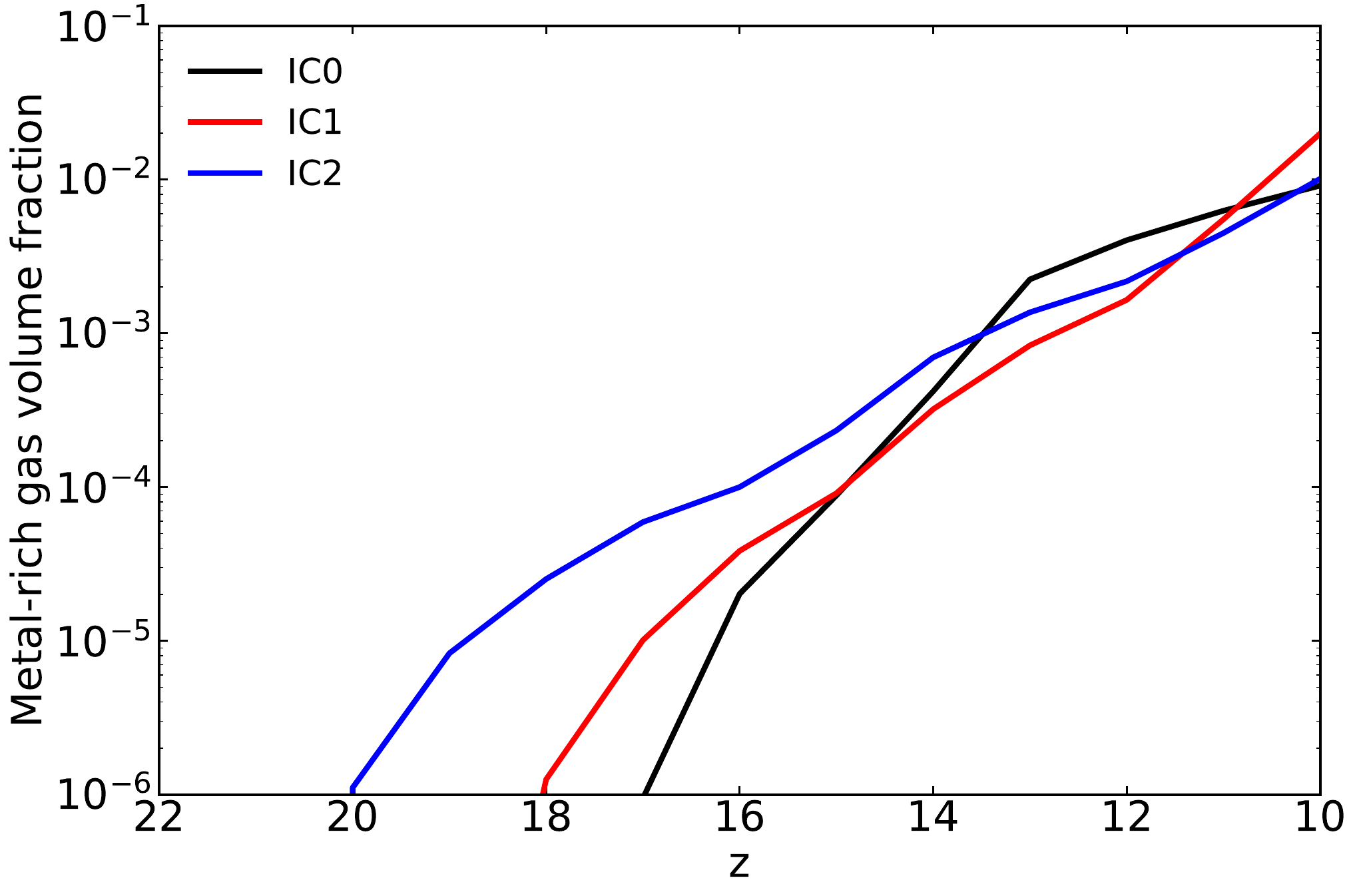}
 \caption{The time evolution of the metal-rich gas ($Z/Z_{\odot}>10^{-4}$) volume filling factor from $z=20$ to $z=10$.}
 \label{fig:std_metal_filling}
\end{figure}

\subsection{Star formation}\label{sec:rsf}

Having presented the qualitative evolution of gas properties in the previous section, we now turn to a quantitative analysis of the stellar populations and star formation activity.

Figure \ref{fig:std_sfr} shows the time evolution of the cumulative stellar mass (upper panel) and star formation rate density (SFRD, bottom panel) for Pop III and Pop II stars across simulations with three different ICs. The upper panel demonstrates that all three ICs reproduce a similar total amount of Pop III and Pop II stars by the end of the simulations ($z=10$). The total mass of Pop III stars is consistent across the runs, reaching around $3\times 10^4\,{\rm M_{\odot}}$ at $z=10$. For Pop II stars, the final stellar masses are generally around several $10^5\,{\rm M_{\odot}}$. 

From the figure, we can see that the IC0 simulation produces approximately half the Pop II stellar mass of the other two simulations. Since Pop II stars preferentially form in massive halos, we investigate the gas state in the two most massive halos in the IC0 simulation. Figure \ref{fig:std_interaction} shows the distributions of surface density, LW radiation intensity, and the ratio of the LW intensity to the critical threshold required for ${\rm H_2}$ cooling suppression for these two subhalos at $z=10$. The black dashed lines represent the virial radii. The subhalo masses are $\sim6\times10^{7}\,{\rm M_{\odot}}$ and $\sim2.5\times10^{7}\,{\rm M_{\odot}}$ at $z=10$. We find that these two massive subhalos reside in a dense filament (which is not fully shown in the figure) and are spatially close to each other, both belonging to the same large friend-of-friend (FoF) halo. Furthermore, the most massive subhalo does not contain any stars, while the second most massive one exhibits significant star formation. As seen in the LW intensity map, continuous star formation in the secondary subhalo generates a strong LW background that irradiates its adjacent neighbor. The center of the most massive subhalo receives non-negligible LW irradiation, which suppresses ${\rm H_2}$ formation and consequently prevents star formation in its center. To further quantify this effect, we compare the $J_{21}/n$ with the critical threshold required to suppress ${\rm H_2}$ cooling, adopting the criterion from \citet{oh2002}. As shown in the rightmost panel, the ratio of $J_{21}/n$ to the critical value exceeds unity by more than two orders of magnitude in the central region of the most massive subhalo. This confirms that the strong external irradiation dissociates ${\rm H_2}$ efficiently, preventing the gas from cooling and collapsing into stars.

Since Pop II stars preferentially form in more massive halos compared to Pop III stars \citep[e.g.][]{yoshida08}, the suppression of star formation in the most massive subhalo of the IC0 simulation directly leads to the slightly lower total Pop II mass observed. In contrast, Pop III stars mainly form in minihalos, so their total amount is not affected significantly by this specific halo-halo interaction. We note that stochastic run-to-run variation in the star formation and feedback model may also contribute to the scatter between different initial conditions (see Appendix \ref{appx:random}).

As shown in the bottom panel of Figure \ref{fig:std_sfr}, for Pop III stars, after the first star forms, the SFRD increases and eventually plateaus at $\sim10^{-4.5}\,{\rm M_{\odot}\,yr^{-1}\, cMpc^{-3}}$ after $z\simeq16$, although with significant stochastic variation. This value is close to that reported in \citet{wise12}, as well as other Pop III cosmological simulations or semi-analytic models \citep[e.g.,][etc.]{jaacks18}. The SFRD of Pop II stars increases with decreasing redshift. The values in the three simulations are similar to each other and consistent with \citet{wise12}, although the IC0 and IC1 simulations exhibit larger variations.

\subsection{Gas properties}\label{sec:rgp}

Figure \ref{fig:std_gas_phase} presents the gas number density-gas temperature ($n-T$) phase diagrams for all gas, metal-poor gas, and metal-rich gas across three simulations with different ICs. We define metal-poor gas as gas with metallicity $Z<10^{-4}\,{\rm Z_{\odot}}$, and metal-rich gas as that with $Z\ge10^{-4}\ {\rm Z_{\odot}}$. The first row displays the total gas component. In all three simulations, the distribution exhibits a similar morphology, which can be categorized into three distinct phases: 1) hot diffuse gas ($n\lesssim1\ {\rm cm^{-3}}$, $T\gtrsim10^4\ {\rm K}$), generated by shock-heating from SN feedback; 2) cold diffuse gas ($n\gtrsim10^{-2}\ {\rm cm^{-3}}$, $T\lesssim10^4\ {\rm K}$), corresponding to the IGM cooling adiabatically due to cosmic expansion; 3) cold dense gas ($n\gtrsim\ 1\ {\rm cm^{-3}}$, $T\lesssim10^4\ {\rm K}$), which represents the primary fuel for star formation.

To disentangle the thermal properties of chemically distinct regions, we separately plot the metal-rich and metal-poor gas in the second and third rows of Figure \ref{fig:std_gas_phase}. Comparing the metal-poor component with the total gas distribution reveals that metal-poor gas dominates the cold diffuse IGM phase. This is expected, as the large-scale IGM remains largely pristine, consistent with the metallicity maps shown in Figure \ref{fig:std_dis}. Notably, a population of hot, diffuse metal-poor gas is also present. This indicates that SN outflows sweep up and shock-heat the surrounding primordial gas, without immediately mixing metals into these swept-up shells. In the cold dense regime, the metal-poor gas follows a relatively tight sequence where temperature decreases with increasing density. This trend arises because cooling in this regime relies on molecular hydrogen (${\rm H_2}$); since the ${\rm H_2}$ fraction correlates with gas density, the equilibrium temperature becomes density-dependent \citep[e.g.,][]{glover08}. We also identify a distinct branch of dense gas ($n>1\,{\rm cm^{-3}}$) with high temperatures ($\sim 10^4\,{\rm K}$) in the IC2 simulation. We attribute this feature to strong local LW radiation fields from nearby star-forming regions, which dissociate ${\rm H_{2}}$ and suppress cooling.

The second row of Figure \ref{fig:std_gas_phase} illustrates the properties of the metal-rich gas. Compared to the metal-poor component, metal-rich gas is preferentially found in the hot diffuse phase (filling the SN bubbles) rather than the cold diffuse IGM. In the cold dense regime, the metal-rich gas exhibits a much broader scatter in the $n-T$ plane. This dispersion arises because metal-line cooling is highly efficient; gas can cool rapidly before it has time to collapse to higher densities, breaking the tight density-temperature correlation seen in primordial gas. The gas in the IC2 simulation extends to the highest densities, reflecting the deeper potential well of the most massive halo formed in that run. Overall, despite the varying IC, the thermodynamic properties of the gas phases are remarkably consistent across all three simulations.

Finally, we examine the time evolution of the metal-enriched gas volume fraction in Figure \ref{fig:std_metal_filling}. Consistent with Figure \ref{fig:std_dis}, the IC2 simulation shows the earliest pollution, while IC0 is enriched later. However, at $z=10$, the volume filling factor of metal-rich gas converges to a similar value of $\sim 1\%$ in all simulations, though with scatter at the factor-of-a-few level (see Appendix \ref{appx:random} for a detailed discussion). This suggests that, on cosmological scales, the signatures of the precise timing of the first Pop III supernovae are largely washed out at $z\sim10$. 

This filling factor is systematically higher than the value reported in previous cosmological simulations (e.g. $\sim 10^{-3}$ in \citealt{jaacks18}). This difference likely reflects a combination of our higher resolution, different feedback implementation, and the inclusion of an explicit turbulent metal mixing model. A controlled comparison (same random seed) with mixing disabled shows a factor of $\sim2$ reduction in the filling factor ($0.91\%$ vs $0.46\%$), indicating that the mixing model contributes to but does not dominate the difference.

\section{Resolution dependence}\label{sec:rd}

In this section, we investigate the numerical convergence and robustness of the subgrid model introduced in Section \ref{sec:method} by varying the simulation resolution. We adopt the IC0 simulation as our fiducial run and perform two additional runs with identical initial conditions but different mass resolutions: a high-resolution run (HRes) and a low-resolution run (LRes). The target gas mass resolution is $\sim350\,{\rm M_{\odot}}$ in the HRes simulation and $\sim 4500\,{\rm M_{\odot}}$ in the LRes simulation, compared to $\sim 1000\,{\rm M_{\odot}}$ in the fiducial run. However, the minimum gas cell size is similar in all runs, as shown in the following results. We mainly present the impact of resolution on the star formation history. The resolution dependence of the gas phase properties in attached in Appendix~\ref{appx:gp}.

\begin{figure}
 \subfigure{\includegraphics[width=1\linewidth]{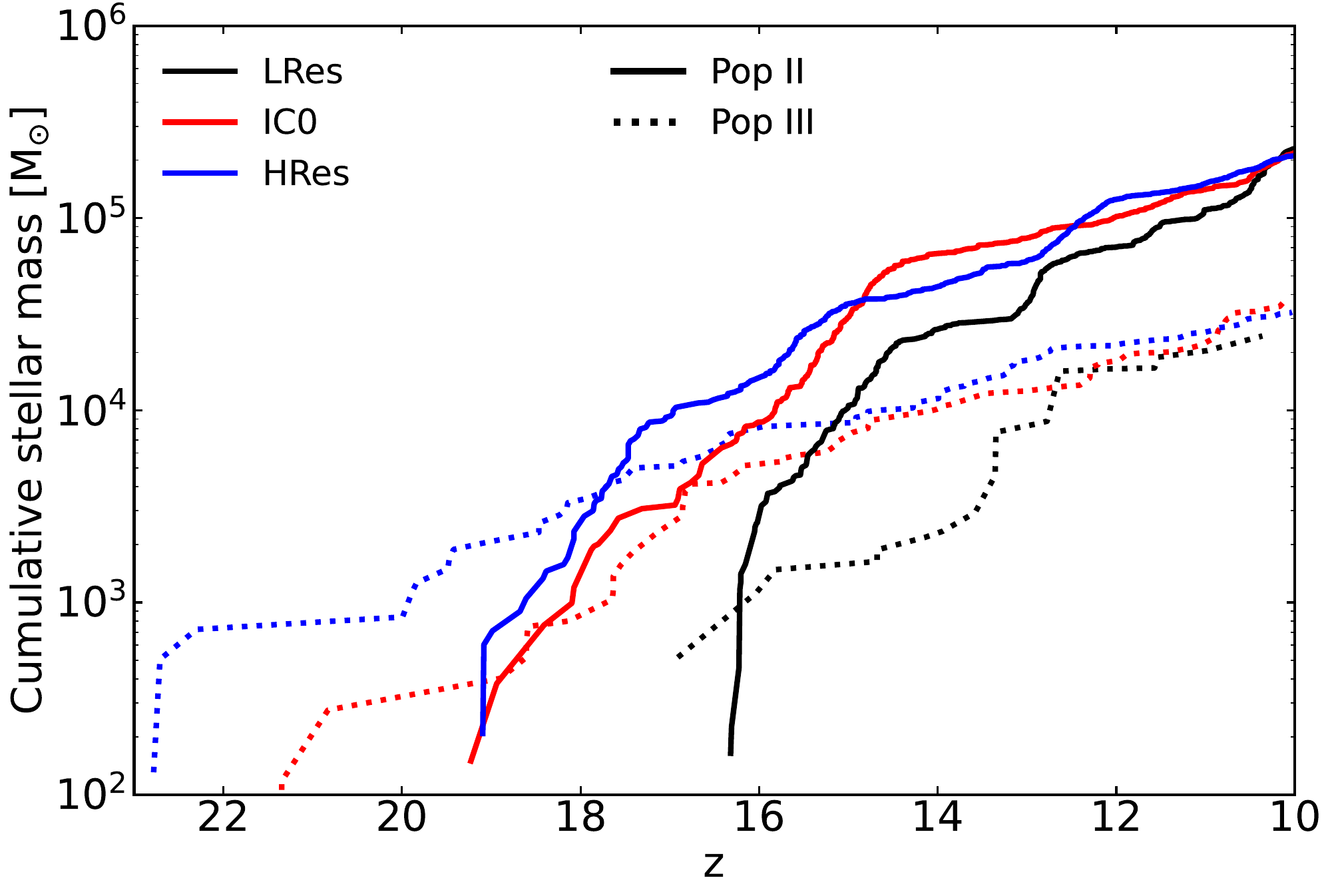}}\\
 \subfigure{\includegraphics[width=1\linewidth]{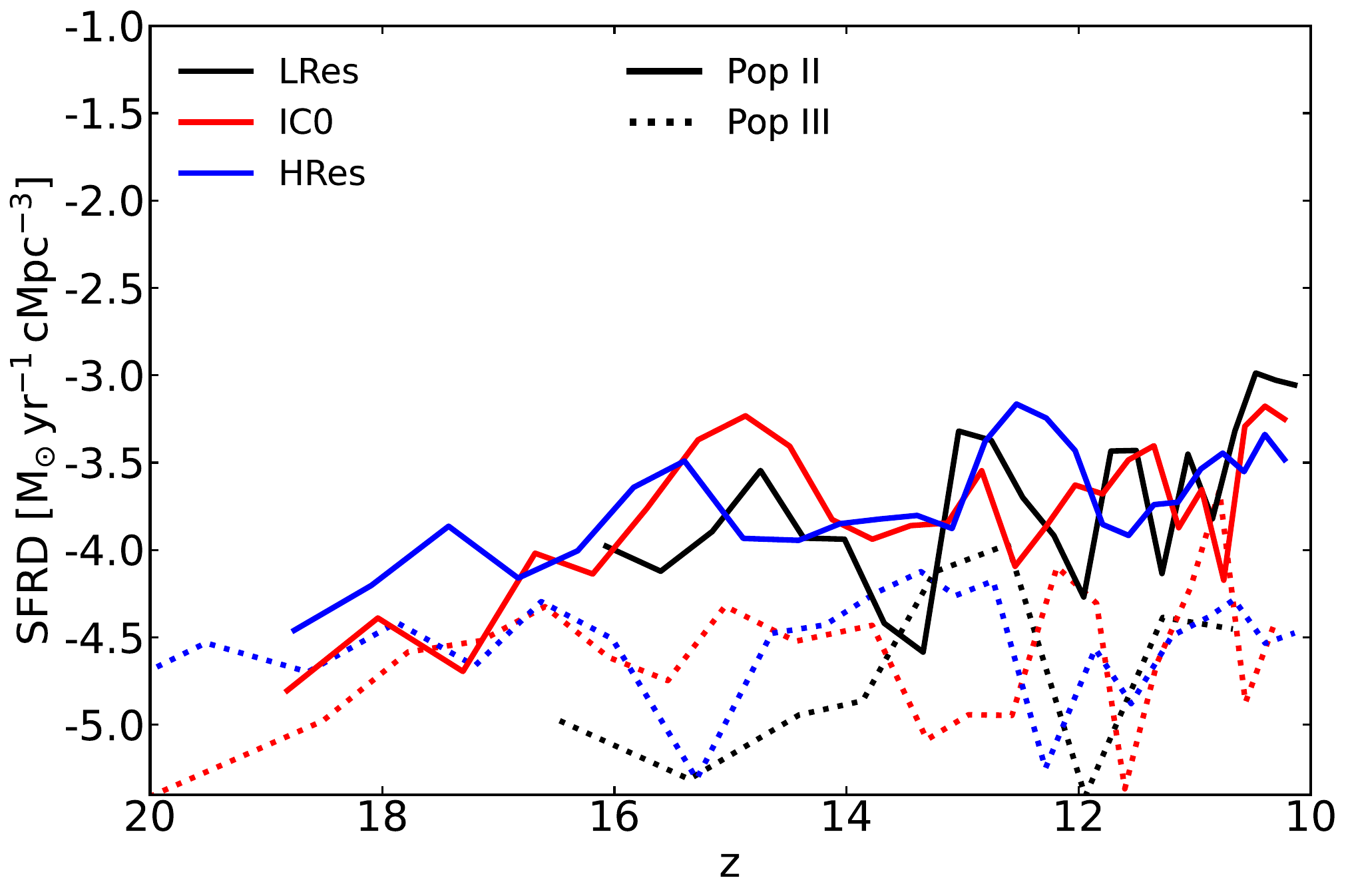}}
 \caption{{\it Upper:} The cumulative Pop III and Pop II stars in the simulations with IC0 but with different resolutions. {\it Bottom:} The Pop III and Pop II star formation rate density (SFRD) from $z=20$ to $z=10$ in three simulations.}
 \label{fig:res_sfr}
\end{figure}

Figure \ref{fig:res_sfr} displays the time evolution of the cumulative stellar mass (top panel) and the star formation rate density (SFRD, bottom panel) for both Pop III and Pop II populations across simulations with varying resolutions. As shown in the top panel, the cumulative Pop III stellar mass is remarkably consistent between the HRes and IC0 simulations. The first Pop III stars in these high-resolution runs form at $z\sim23$ and $z\sim21$, respectively. In contrast, star formation in the LRes simulation is delayed, with the cumulative mass beginning to rise only at $z\sim17$. Despite this initial discrepancy, the total cumulative stellar mass at $z=10$ is similar among the three simulations, although the detailed timing of star formation remains strongly resolution-dependent. This result implies that the total reservoir of gas capable of collapsing and forming stars is relatively robust to resolution limits at $z=10$, but the onset of structure formation is resolution-dependent.

For Pop II stars, we observe a similar trend. The evolution of the cumulative Pop II stellar mass shows even better agreement between the HRes and IC0 simulations compared to the LRes run. Furthermore, the resolution convergence appears to be faster for Pop II stars than for Pop III. Since Pop II formation is triggered by metal enrichment from prior Pop III supernovae, the delay in Pop III formation in the LRes run naturally leads to a delayed onset of Pop II stars ($z\sim16$, compared to $z\sim19$ in the high-resolution runs). Nevertheless, similar to the Pop III case, the total Pop II stellar mass at $z=10$ is largely insensitive to resolution.

The bottom panel of Figure \ref{fig:res_sfr} shows the evolution of the SFRD. At $z>16$, the SFRD is effectively zero in the LRes simulation due to the delayed onset of star formation. However, at $z<14$, the SFRD in all simulations becomes comparable, exhibiting similar stochastic fluctuations. These results demonstrate that the IC0 and HRes simulations produce consistent star formation histories, whereas the LRes simulation suffers from a systematic delay due to its inability to resolve the earliest minihalos. This indicates that our fiducial resolution (IC0) has achieved convergence for the physical quantities of interest in this work.


\begin{figure}
\includegraphics[width=1\linewidth]{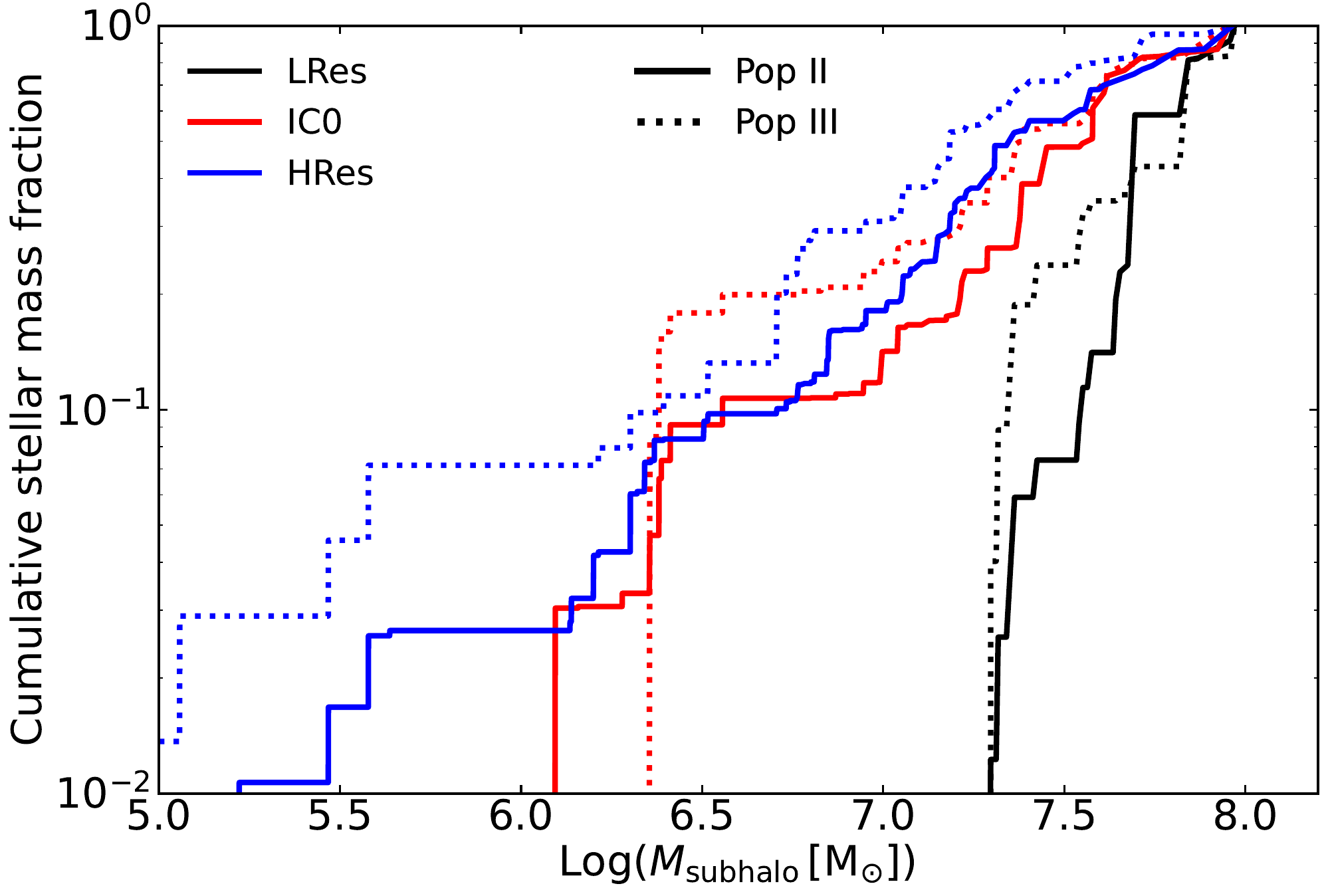}
 \caption{The cumulative stellar mass fraction as a function of subhalo mass for simulations with three different resolutions (LRes, IC0 and HRes). Solid and dashed lines correspond to the contributions from Population II and Population III stars, respectively. The plot demonstrates that $\sim90\%$ of stellar mass, for both Pop II and Pop III, is hosted by subhalos with mass $M_{\rm subhalo}\gtrsim 10^{6.5}\,{\rm M_{\odot}}$. All data are taken from the z=10 snapshot.}
 \label{fig:res_fstar}
\end{figure}

To better understand the origin of the discrepancy in the LRes simulation, we examine the cumulative distribution of Pop III and Pop II stellar mass as a function of host subhalo mass in Figure \ref{fig:res_fstar}. In the HRes simulation, approximately $50\%$ of both Pop III and Pop II stars reside in subhalos with $M_{\rm subhalo} \lesssim 10^{7.3}\, {\rm M_{\odot}}$. Notably, only $\sim10\%$ of the stellar mass is found in subhalos with $M_{\rm subhalo}\lesssim10^{6.5}\,{\rm M_{\odot}}$. This implies that a simulation must resolve gas condensation within subhalos of mass $M_{\rm subhalo}\sim10^{6.5}\,{\rm M_{\odot}}$ to capture the majority ($90\%$) of the star formation history. 

In the LRes simulation, the minimum mass of a star-hosting subhalo is $\sim 10^{7.3}\,{\rm M_{\odot}}$. This indicates that the density threshold for Pop III star formation is only reached in halos above this mass limit, effectively missing the contribution from lower-mass minihalos capable of forming stars. Conversely, in the IC0 simulation, the minimum host mass is $\sim 10^{6.1}\,{\rm M_{\odot}}$, sufficient to capture $>90\%$ of the total star formation, explaining the strong agreement between the IC0 and HRes star formation histories. Interestingly, in the IC0 simulation, we observe that the minimum subhalo mass hosting Pop II stars is slightly lower than that of Pop III hosts. This implies that the metals triggering Pop II formation in these low-mass systems do not originate from in-situ Pop III supernovae within the same subhalo. Instead, they are likely enriched externally by supernovae in neighboring subhalos via inter-halo metal transport (external enrichment).

Furthermore, Pop II stars generally reside in more massive subhalos than Pop III stars. This trend is expected: Pop III stars predominantly form in low-mass minihalos with shallow potential wells, allowing supernova ejecta to escape efficiently (blow-away) without enriching the host halo. In contrast, massive subhalos can retain metals more effectively and are enriched via the accretion of metal-rich outflows from nearby minihalos or through the hierarchical merger of enriched progenitors.

We note that the resolution tests presented here cover a relatively modest range of gas mass resolutions (a factor of $\sim$12 between LRes and HRes). While our results demonstrate convergence within this range, extrapolation to significantly different resolutions should be made with caution.

\section{Summary}\label{sec:summary}

In this work, we presented the SPARK framework, which is a new Pop III + Pop II subgrid framework implemented in the moving-mesh code {\sc arepo}. The model includes {\bf (1)} star-formation criteria for primordial and metal-enriched gas, {\bf (2)} IMF-weighted stellar evolution with Poisson-sampled SN injection, {\bf (3)} stellar winds for massive Pop II stars, {\bf (4)} approximate ionizing and LW radiation transport, {\bf (5)} non-equilibrium primordial chemistry, {\bf (6)} metal-line cooling tables coupled to stellar radiation, and {\bf (7)} an adaptive, convergent Jeans-based refinement strategy. This combination of modules forms a self-consistent framework for Pop III + Pop II transition studies in cosmological volumes with {\sc arepo} code.

Using this framework, we performed multiple $1\,c{\rm Mpc}/h$ box simulations with different initial conditions and resolutions. Our main findings are:

\begin{itemize}

    \item The model reproduces the Pop II star formation rate density  consistent with the results of \citet{wise12} across all initial conditions, with minor variations mainly arising from local halo–halo interactions and LW irradiation (Figure \ref{fig:std_sfr} and \ref{fig:std_interaction}).

    \item The volume filling factor of metal-enriched gas reaches values of order $\sim1\%$ at $z = 10$, with moderate scatter across different runs (Appendix \ref{appx:random}), suggesting that our feedback model drives bubble expansion efficiently and robustly (\ref{fig:std_metal_filling} and \ref{fig:res_metal_filling}).

    \item Resolution tests show convergence once halos with mass $M_{\rm subhalo}>10^{6.5}\,{\rm M_\odot}$ are resolved, indicating that capturing the dense gas in these systems is sufficient for converged Pop III/II star-formation histories. Under our model, this corresponds to 500-1000 DM particles for $M_{\rm subhalo}\sim 10^{6.5}\,{\rm M_{\odot}}$ subhalo (Figure \ref{fig:res_sfr} and  \ref{fig:res_fstar}).

    \item The total stellar mass formed by $z=10$ is largely insensitive to IC and resolution, suggesting that early star formation is primarily regulated by the amount of gas able to collapse (Figure \ref{fig:std_sfr} and \ref{fig:res_sfr}).
\end{itemize}

A key advantage of our framework is its computational efficiency. By employing an approximate RT scheme and a mesh refinement strategy, the model strikes a balance between physical fidelity and cost: a single $1\, c{\rm Mpc}/h$ fiducial run from $z=127$ to $z=10$ requires only $\sim 10^4$ CPU hours, typically using 640–896 MPI tasks on Intel Xeon Gold nodes. This tractable computational cost, significantly lower than full-RT calculations, enables the exploration of multiple initial conditions, rigorous convergence testing, and future large-scale parameter studies or zoom-in simulations.

Some simplifications remain in the current model. The current radiation model includes two bands (LW and ionizing), without explicit treatment of X-ray photons or spectral separation of H- and He-ionizing radiation. The ionizing region is also estimated using a spherical gas distribution around each source. Despite these approximations, the model successfully reproduces key global observables, indicating that the dominant physical processes are captured at the scales resolved here.

In future work, we will extend the radiation model and apply this framework to investigate Pop III IMF variations, X-ray feedback, and the environmental dependence of early metal enrichment.

\begin{acknowledgments}
We thank the anonymous referee for a constructive report that improved this paper. We thanks Oliver Zier, Zhiyuan Yao, Zhi Li, Yizhou Liu, and Hang Yang for thorough comments, discussions and suggestions. We thank Volker Springel for kindly authorizing us the \textsc{arepo} code. We acknowledge support from the National Natural Science Foundation of China (Grant No.
12588202) and the National Key Research and Development Program of China (Grant No. 2023YFB3002500).
H.H. is supported by the NSFC Grant Nos. 12503012.
\end{acknowledgments}

\appendix

\section{Dependence on Star Formation Parameters}\label{appx:sfcrit}

\begin{figure}
\includegraphics[width=1\linewidth]{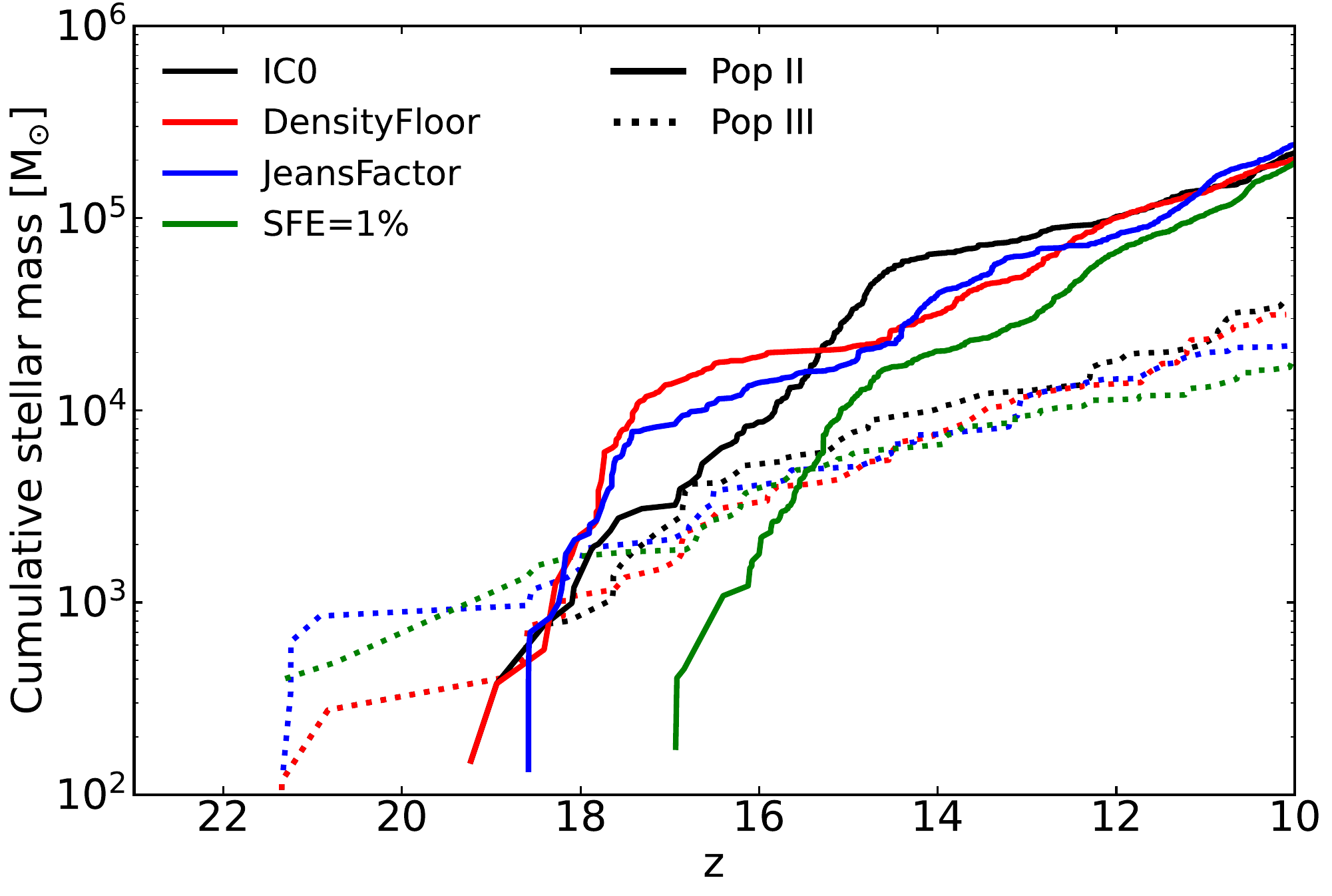}
 \caption{The cumulative Pop III and Pop II stars in the simulation with varied star formation parameters: DensityFloor ($n_{\rm th}=\max(10\,{\rm cm^{-3}}, 0.5\, n_{\rm Jeans})$), JeansFactor ($n_{\rm th}=\max(1\,{\rm cm^{-3}}, 0.8\, n_{\rm Jeans})$), and SFE=$1\%$.}
 \label{fig:sf_csm}
\end{figure}

\begin{figure}
\includegraphics[width=1\linewidth]{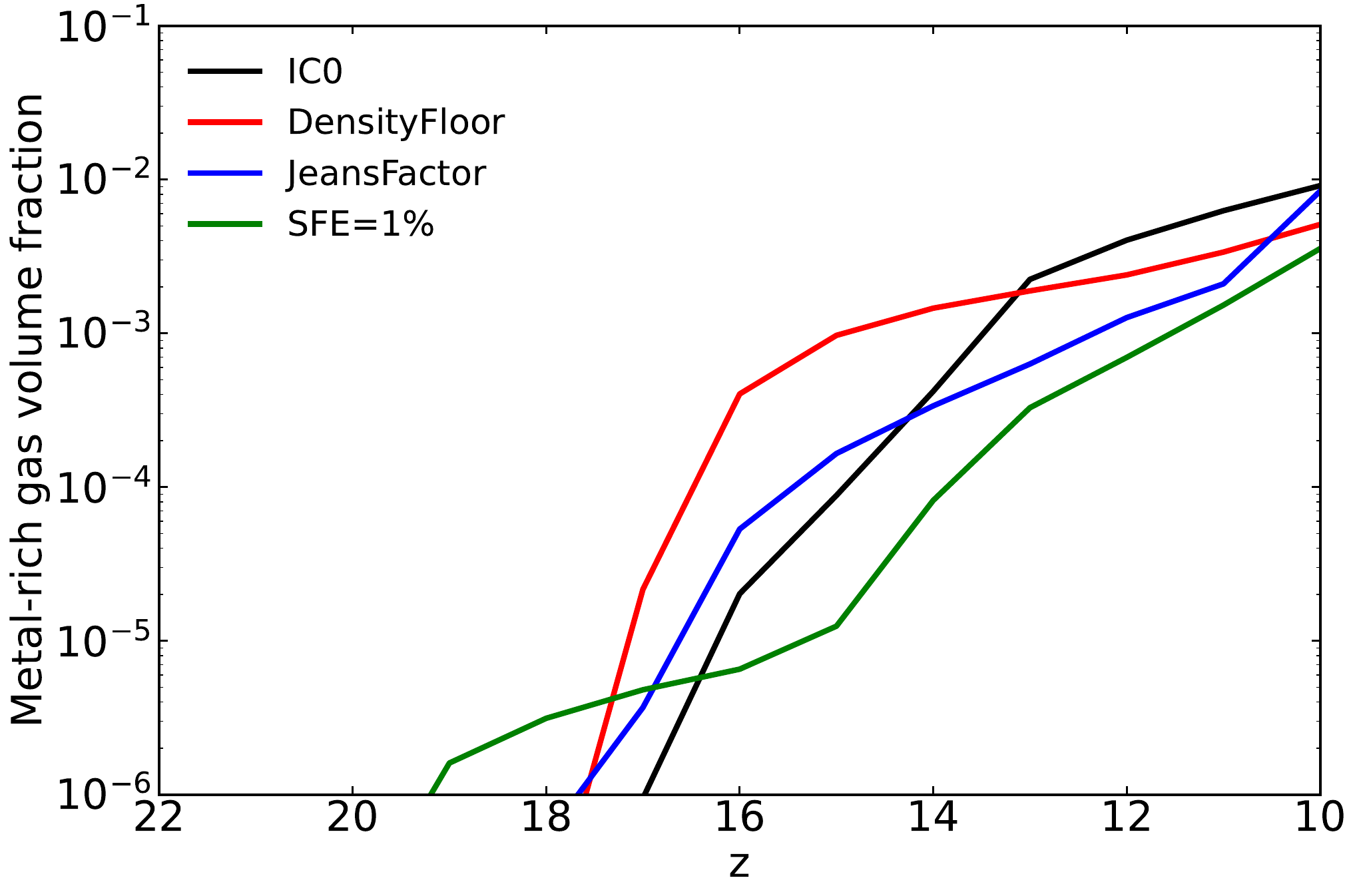}
 \caption{Same as Figure~\ref{fig:sf_csm} but for the time evolution of the metal-rich gas ($Z/Z_{\odot}>10^{-4}$) volume filling factor from $z=20$ to $z=10$.}
 \label{fig:sf_filling}
\end{figure}

To assess the robustness of our results against the uncertainties in subgrid parameters, we performed three additional simulations based on the IC0 initial conditions. We varied the Jeans prefactor from 0.5 to 0.8 (labeled JeansFactor), the star formation density threshold ($n_{\rm th}$=$\max(10\,{\rm cm^{-3}}, 0.5\, n_{\rm Jeans})$) (labeled DensityFloor), and the SFE to $1\%$ (labeled SFE=$1\%$). Refer to Section \ref{sec:pop3sf} for details on the fiducial model.

Figure \ref{fig:sf_csm} presents the cumulative stellar mass evolution for these variations compared to the fiducial run (IC0). The DensityFloor and JeansFactor simulations yield total stellar masses at $z = 10$ consistent with the fiducial run, with differences comparable to the stochastic run-to-run variation documented in Appendix \ref{appx:random}. Conversely, the SFE=1\% model exhibits a noticeable delay in the onset of Pop II star formation, although the total stellar mass converges toward the fiducial result by $z=10$. For Pop III stars, the total stellar mass in the low-SFE run is approximately $2/3$ of that in the fiducial run. This suggests that the impact of the local SFE parameter on the global stellar mass is sub-linear.

Figure \ref{fig:sf_filling} illustrates the impact of these parameters on the volume fraction of metal-rich gas. The volume filling factors in the DensityFloor and JeansFactor simulations show different evolutionary histories from the fiducial run at early times. These differences arise from the stochastic nature of the star formation and feedback modeling for the first few Pop III stars, which affects the onset of Pop II star formation and the subsequent growth of metal-enriched bubbles. However, by $z=10$, the filling factors converge to values within the stochastic run-to-run variation documented in Appendix \ref{appx:random}. The SFE=1\% simulation displays a significant delay in metal enrichment. At $z=10$, the volume filling factor in this run is approximately half that of the fiducial simulation. This reduction is likely a timing effect, as the lower star formation efficiency delays the overall star formation history and the subsequent supernova feedback, leaving less time for the metal-enriched bubbles to expand by $z = 10$.

\section{Chemical network}\label{appx:reaction}

The chemical network is selected from \citet{katz96, abel97, glover08}, we choose some key atomic, molecular and photoionization/photodissociation processes. The full set of reactions is listed in Table \ref{tab:network}. The chemical network is solved by an in-house implementation of a backward Euler scheme combined with a Jacobian-Free Newton-Krylov method within Arepo. 

\begin{table*}
\caption{The chemical network for primordial gas used in this work. Reaction rates are taken from the references listed in the table.}
\centering
\begin{tabular}{c}
\hline
Selected processes from \citet{katz96} \\
\hline
${\rm H+e^{-}\rightarrow H^{+}+2e^-}$\\
${\rm H^++e^{-}\rightarrow H+\gamma}$\\
${\rm He+e^{-}\rightarrow He^++2e^-}$\\
${\rm He^++e^{-}\rightarrow He+\gamma}$\\
${\rm He^{+}+e^{-}\rightarrow He^{++}+2e^+}$\\
${\rm He^{++}+e^{-}\rightarrow He^{+}+\gamma}$\\

\hline
Selected processes from \citet{abel97} \\
\hline
${\rm H+e^{-}\rightarrow H^{-}+\gamma}$\\
${\rm H+H^-\rightarrow H_2+e^-}$\\
${\rm H_2+H^+\rightarrow H_2^++H}$\\
${\rm H_2+e^-\rightarrow 2H+e^-}$\\
${\rm H^-+e^-\rightarrow H+2e^-}$\\
${\rm H^++H^-\rightarrow 2H}$\\
\hline
Selected processes from \citet{glover08}\\
\hline
${\rm H_2+H\rightarrow 3H}$\\
${\rm H+H^+\rightarrow H_2^++\gamma}$\\
${\rm H_2^++H\rightarrow H_2+H^+}$\\
${\rm H+H\rightarrow H^++H+e^-}$\\
${\rm H^-+H\rightarrow 2H+e^-}$\\
\hline
Selected photoionization/photodissociation processes from \citet{abel97}\\
\hline
${\rm H+\gamma\rightarrow H^++e^-}$\\
${\rm He+\gamma\rightarrow He^++e^-}$\\
${\rm He^++\gamma\rightarrow He^{++}+e^-}$\\
${\rm H^-+\gamma\rightarrow H+e^-}$\\
${\rm H_2+\gamma\rightarrow H_2^++e^-}$\\
${\rm H_2^++\gamma\rightarrow H^++H}$\\
${\rm H_2^++\gamma\rightarrow 2H^++e^-}$\\
${\rm H_2+\gamma\rightarrow H_2^*\rightarrow 2H}$\\
${\rm H_2+\gamma\rightarrow 2H}$\\
\hline
\end{tabular}
\label{tab:network}
\end{table*}

\begin{figure}
\includegraphics[width=\linewidth]{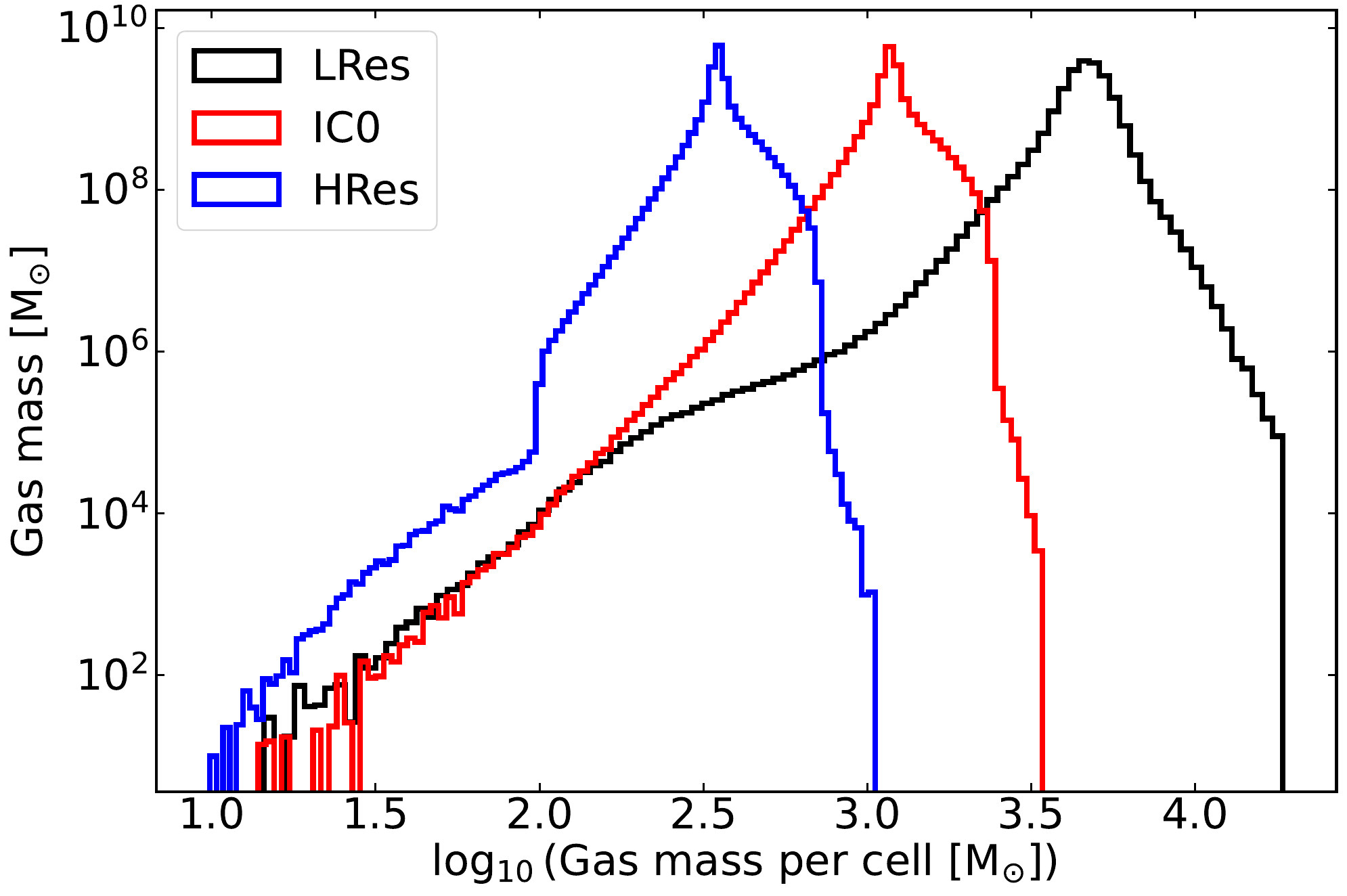}
 \caption{The gas cell mass distribution in the simulations with IC0 initial condition but different resolutions.}
 \label{fig:res_gas_cell}
\end{figure}

\begin{figure*}
 \includegraphics[width=\linewidth]{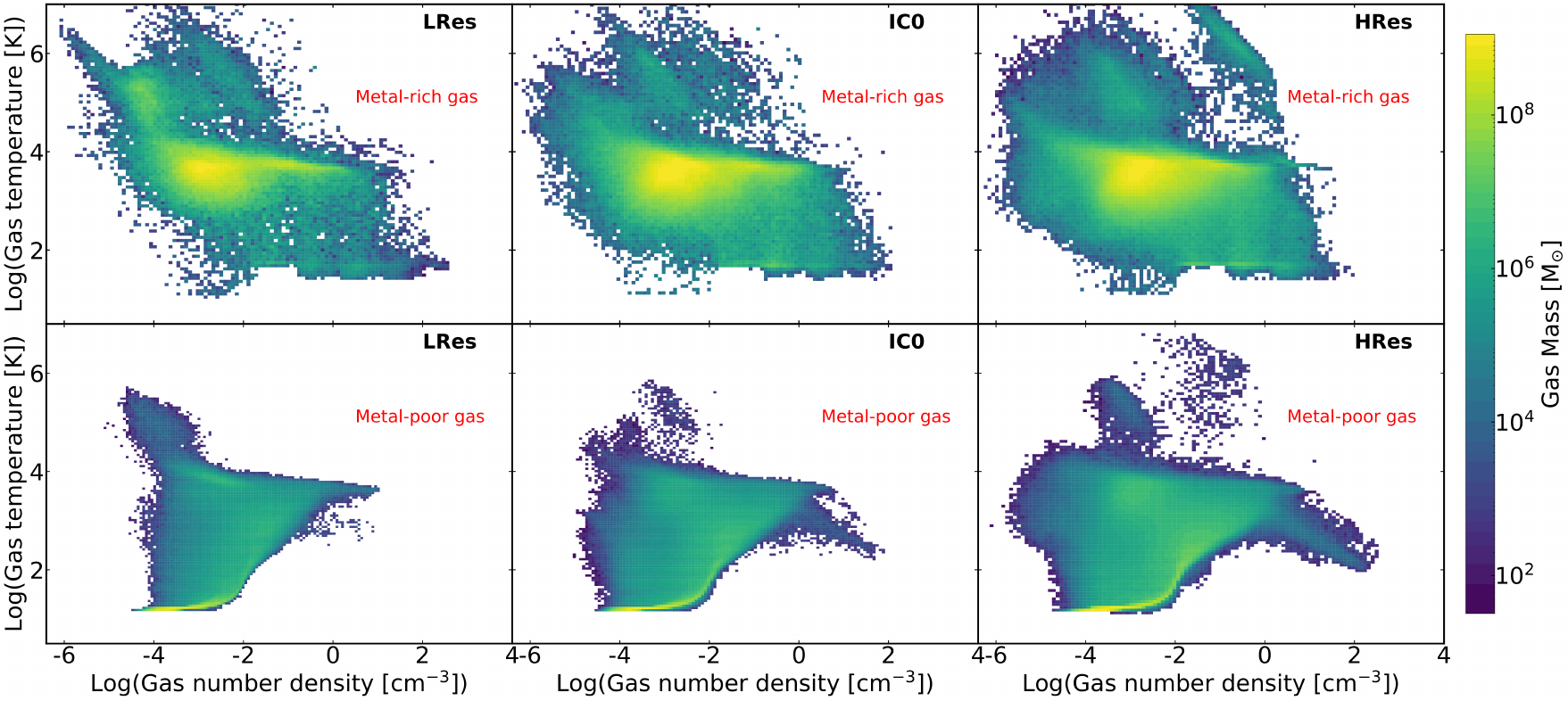}
 \caption{Similar to Figure \ref{fig:std_gas_phase}. The gas phase diagram of number density and temperature for metal-rich gas ({\it upper}, defined by $Z/Z_{\odot}>10^{-4}$) and metal-poor gas({\it bottom}, defined by $Z/Z_{\odot}\le10^{-4}$) at z=10. Each column represents the results from the same simulation.}
 \label{fig:res_gas_phase}
\end{figure*}

\begin{figure}
\includegraphics[width=\linewidth]{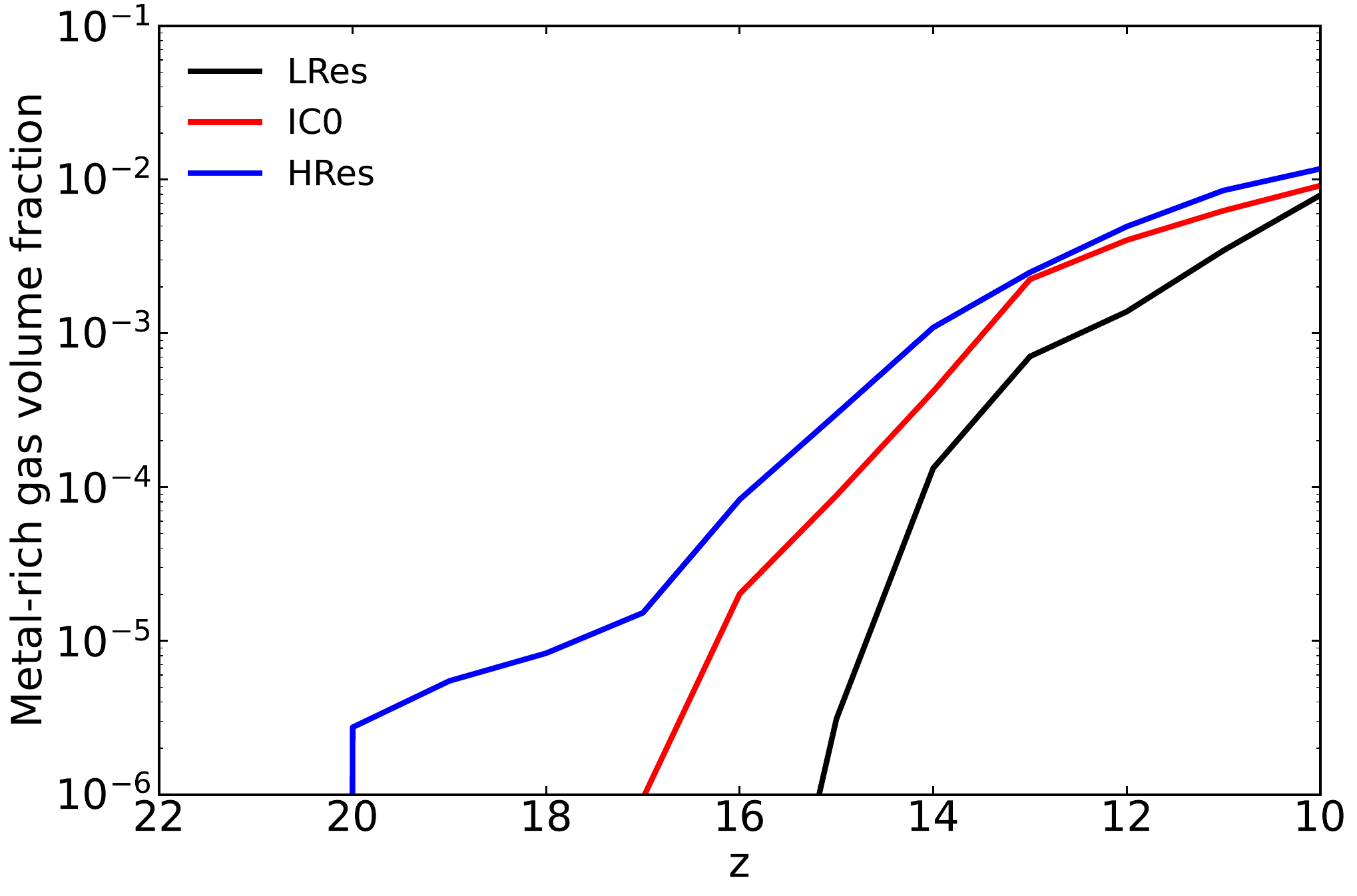}
 \caption{Similar to Figure \ref{fig:std_metal_filling}, but for  the simulations with different resolution at $z=10$.}
 \label{fig:res_metal_filling}
\end{figure}

\section{Gas properties in the simulations with different resolutions}\label{appx:gp}

We examine the distribution of gas cell masses for the three simulations with different resolutions, as shown in Figure \ref{fig:res_gas_cell}. The distribution exhibits a skewed shape with a tail extending towards low masses. The peak of the distribution corresponds to the target mass resolution of each simulation. The low-mass tail arises from Jeans refinement. The distribution of gas cells with mass $\lesssim100\,{\rm M_{\odot}}$ shows a similar shape across IC0 and LRes, while the HRes simulation contains significantly more mass in this regime. We find that this excess is predominantly composed of hot gas ($T\gtrsim10^4\,{\rm K}$) from the resolved interiors of supernova remnants, which are protected from de-refinement by our mesh management strategy (Section \ref{sec:refinement}). We note that the similarity of the cell mass distribution in the star-forming regime does not imply convergence by construction: the LRes simulation, despite a comparable cold gas cell mass distribution, exhibits a significant delay in star formation due to its inability to resolve low-mass minihalos in the DM component. An additional test with enhanced DM resolution but fiducial gas resolution confirms that both components must be adequately resolved for star formation to proceed in low-mass halos.

We investigate the thermodynamic state of the gas in the three simulations. Figure \ref{fig:res_gas_phase} shows the $n-T$ phase diagrams for metal-rich and metal-poor gas separately. For the metal-rich gas, the LRes simulation successfully captures the dense, cold gas phase. Due to the effective cooling provided by metals, gas can undergo rapid collapse and cooling, allowing the LRes simulation to better resolve the conditions required for Pop II star formation. We also observe a population of hot, dense, metal-enriched gas ($T>10^4\,{\rm K}$ and $n\sim10^{-1}-10^{1}\, {\rm cm^{-3}}$) in the HRes simulation, which corresponds to the resolved interiors of metal-rich SN bubbles.

The bottom panel of Figure \ref{fig:res_gas_phase} shows the phase diagram for metal-poor gas. In contrast to the metal-rich case, the LRes simulation lacks a stable, high-density, low-temperature branch due to insufficient resolution. Since the density criterion for Pop III star formation is set to $\min(0.5\,n_{\rm Jeans}, 1\,{\rm cm^{-3}})$, the LRes simulation only marginally resolves the conditions for Pop III star formation. Comparing the HRes and IC0 simulations, we find that the cold dense gas is better resolved in the HRes run, with a larger reservoir of gas residing in the cold, dense phase. In addition, a distinct population of gas with high temperature and high density ($T>10^4\,{\rm K}$ and $n\sim10^{-2}-1\, {\rm cm^{-3}}$) appears in the HRes simulation. This gas corresponds to the hot, shocked interiors of SN bubbles, which also appear in a previous simulation study \citep{bromm03firstSN}. Due to its higher resolution, the HRes simulation can better resolve the internal structure of these SN remnants compared to IC0.

Figure \ref{fig:res_metal_filling} shows the time evolution of the metal-rich gas volume fraction across the three resolution levels. Similar to the star formation history, the onset of metal enrichment is delayed in the LRes simulation ($z\sim15$) compared to IC0 (z$\sim$20) and HRes ($z\sim23$) due to the delayed collapse of the first minihalos. However, a striking feature is the strong convergence at lower redshifts: at $z=10$, the volume filling factor in all three simulations asymptotes to values of order $\sim1\%$. Since the volume filling factor is directly determined by the size of metal-enriched supernova bubbles, this convergence suggests that our feedback model drives bubble expansion robustly and efficiently across the resolutions considered. 

In summary, while higher resolution naturally captures a broader and more detailed distribution of gas phases (particularly SN bubble interiors), the simulation with our fiducial resolution (IC0) is sufficient to capture the key global star formation history. Conversely, the lower resolution (LRes) fails to capture the early star formation history, primarily because it cannot resolve the dense, primordial gas in minihalos at high redshifts. Therefore, our fiducial resolution represents an optimal balance between computational cost and the ability to resolve key physical processes.

\section{Dependence on stochastic modeling for star formation, stellar feedback, and turbulent metal mixing}\label{appx:random}

\begin{figure}
 \includegraphics[width=\linewidth]{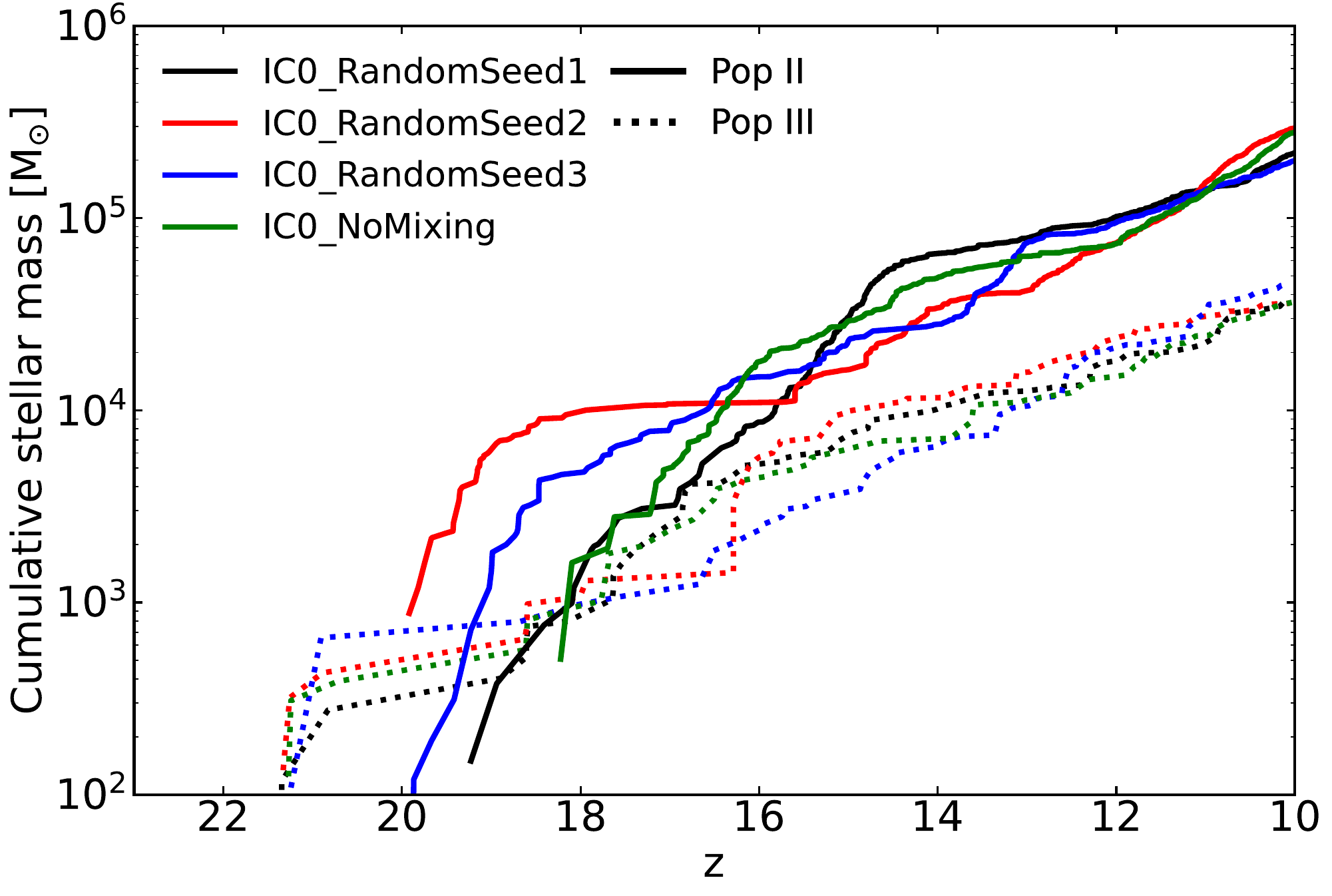}
 \caption{The cumulative Pop~III and Pop~II stellar mass in simulations with IC0 but different random seeds for the stochastic star formation model, as well as a run without turbulent metal mixing. The stellar mass is well converged across all runs at $z\lesssim15$.}
 \label{fig:ic_csm}
\end{figure}

\begin{figure}
 \includegraphics[width=\linewidth]{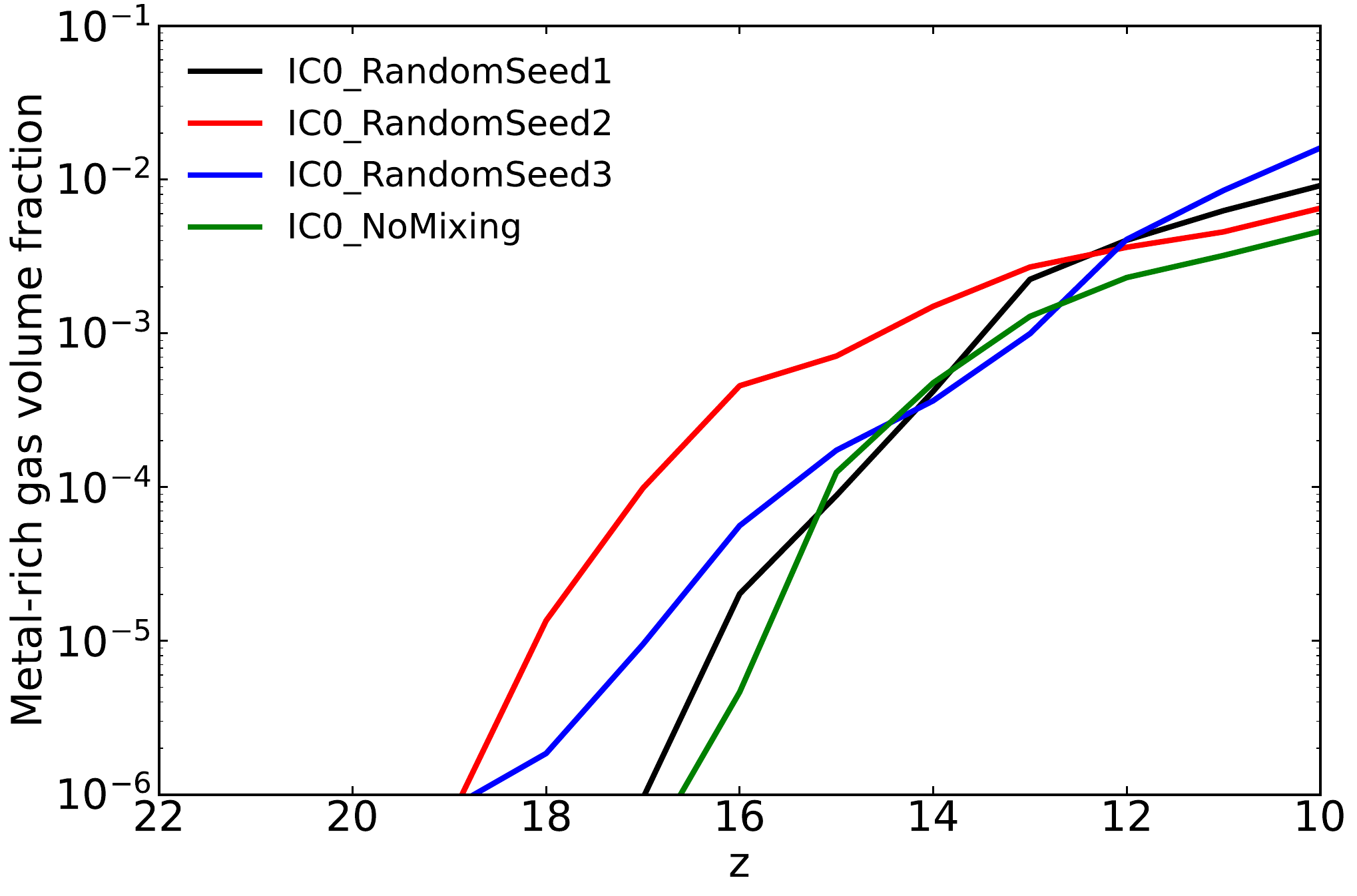}
 \caption{Similar to Figure \ref{fig:std_metal_filling} but for the simulations with the same set of runs as Figure~\ref{fig:ic_csm}. The filling factor shows larger scatter than the stellar mass, ranging from $\sim 0.65\%$ to $\sim 1.6\%$ at $z=10$.}
 \label{fig:ic_metal}
\end{figure}

To assess the sensitivity of our results to the stochastic nature of the star formation model and the turbulent metal mixing prescription, we perform three additional runs with the IC0 initial conditions but different random seeds for the stochastic star formation model, as well as one run with turbulent mixing disabled (C$_{\rm diff}=0$). Such stochastic run-to-run variation is a known feature of cosmological simulations with subgrid models \citep[e.g.,][]{genel19,zhu2025}.

Figure~\ref{fig:ic_csm} shows the cumulative stellar mass for these runs. The total Pop~II and Pop~III stellar masses at $z=10$ are well converged across all runs, confirming that the star formation history is robust to both stochastic variation and the mixing prescription.

Figure~\ref{fig:ic_metal} shows the corresponding volume filling factor of metal-enriched gas. In contrast to the stellar mass, the filling factor exhibits larger scatter, ranging from $\sim 0.65\%$ to $\sim 1.6\%$ across different random seeds at $z=10$. The variation in the onset of metal enrichment across runs reflects the stochastic timing of the first supernovae in different realizations. A controlled comparison using the same random seed with and without turbulent mixing yields $0.91\%$ versus $0.46\%$, indicating a factor of $\sim 2$ sensitivity to the diffusion prescription. The order of magnitude of the filling factor ($\sim 10^{-3}$--$10^{-2}$) is nevertheless robust across all runs.

\bibliography{apssamp}

@ARTICLE{donnan2023,
   author = "Donnan, C. T. and McLeod, D. J. and Dunlop, J. S. and McLure, R. J. and Carnall, A. C. and Begley, R. and others",
   title = "The evolution of the galaxy UV luminosity function at redshifts z ~ 8 - 15 from deep JWST and ground-based near-infrared imaging",
   journal = "Mon. Not. R. Astron. Soc.",
   year = "2023",
   volume = "518",
   pages = "6011-6040",
   doi = "10.1093/mnras/stac3472"
}

@ARTICLE{Weinberger2020,
   author = "Weinberger, Rainer and Springel, Volker and Pakmor, Rüdiger",
   title = "The AREPO Public Code Release",
   journal = "Astrophys. J. Suppl.",
   year = "2020",
   volume = "248",
   pages = "32",
   doi = "10.3847/1538-4365/ab908c"
}

@ARTICLE{Springel2010,
   author = "Springel, Volker",
   title = "E pur si muove: Galilean-invariant cosmological hydrodynamical simulations on a moving mesh",
   journal = "Mon. Not. R. Astron. Soc.",
   year = "2010",
   volume = "401",
   pages = "791-851",
   doi = "10.1111/j.1365-2966.2009.15715.x"
}

@ARTICLE{bromm04,
   author = "Bromm, Volker and Larson, Richard B.",
   title = "The First Stars",
   journal = "Annu. Rev. Astron. Astrophys.",
   year = "2004",
   volume = "42",
   pages = "79-118",
   doi = "10.1146/annurev.astro.42.053102.134034"
}

@ARTICLE{Klessen23,
   author = "Klessen, Ralf S. and Glover, Simon C. O.",
   title = "The First Stars: Formation, Properties, and Impact",
   journal = "Annu. Rev. Astron. Astrophys.",
   year = "2023",
   volume = "61",
   pages = "65-130",
   doi = "10.1146/annurev-astro-071221-053453"
}

@ARTICLE{greif07,
   author = "Greif, Thomas H. and Johnson, Jarrett L. and Bromm, Volker and Klessen, Ralf S.",
   title = "The First Supernova Explosions: Energetics, Feedback, and Chemical Enrichment",
   journal = "Astrophys. J.",
   year = "2007",
   volume = "670",
   pages = "1-14",
   doi = "10.1086/522028"
}

@ARTICLE{wise12,
   author = "Wise, John H. and Turk, Matthew J. and Norman, Michael L. and Abel, Tom",
   title = "The Birth of a Galaxy: Primordial Metal Enrichment and Stellar Populations",
   journal = "Astrophys. J.",
   year = "2012",
   volume = "745",
   pages = "50",
   doi = "10.1088/0004-637X/745/1/50"
}

@ARTICLE{barkana01,
   author = "Barkana, R. and Loeb, A.",
   title = "In the beginning: the first sources of light and the reionization of the universe",
   journal = "Phys. Rep.",
   year = "2001",
   volume = "349",
   pages = "125-238",
   doi = "10.1016/S0370-1573(01)00019-9"
}

@ARTICLE{greif10,
   author = "Greif, Thomas H. and Glover, Simon C. O. and Bromm, Volker and Klessen, Ralf S.",
   title = "The First Galaxies: Chemical Enrichment, Mixing, and Star Formation",
   journal = "Astrophys. J.",
   year = "2010",
   volume = "716",
   pages = "510-520",
   doi = "10.1088/0004-637X/716/1/510"
}

@ARTICLE{wise08,
   author = "Wise, John H. and Abel, Tom",
   title = "Resolving the Formation of Protogalaxies. III. Feedback from the First Stars",
   journal = "Astrophys. J.",
   year = "2008",
   volume = "685",
   pages = "40-56",
   doi = "10.1086/590417"
}

@ARTICLE{jeon14,
   author = "Jeon, Myoungwon and Pawlik, Andreas H. and Bromm, Volker and Milosavljević, Miloš",
   title = "Recovery from Population III supernova explosions and the onset of second-generation star formation",
   journal = "Mon. Not. R. Astron. Soc.",
   year = "2014",
   volume = "444",
   pages = "3288-3300",
   doi = "10.1093/mnras/stu1980"
}

@ARTICLE{ritter12,
   author = "Ritter, Jeremy S. and Safranek-Shrader, Chalence and Gnat, Orly and Milosavljević, Miloš and Bromm, Volker",
   title = "Confined Population III Enrichment and the Prospects for Prompt Second-generation Star Formation",
   journal = "Astrophys. J.",
   year = "2012",
   volume = "761",
   pages = "56",
   doi = "10.1088/0004-637X/761/1/56"
}

@ARTICLE{chiaki19,
   author = "Chiaki, Gen and Wise, John H.",
   title = "Seeding the second star: enrichment from population III, dust evolution, and cloud collapse",
   journal = "Mon. Not. R. Astron. Soc.",
   year = "2019",
   volume = "482",
   pages = "3933-3949",
   doi = "10.1093/mnras/sty2984"
}

@ARTICLE{storck25,
   author = "Storck, Anatole and Katz, Harley and Devriendt, Julien and Slyz, Adrianne and Cadiou, Corentin and others",
   title = "MEGATRON: The environments of Population III stars at Cosmic Dawn and their connection to present day galaxies",
   journal = "arXiv e-prints",
   year = "2025",
   pages = "arXiv:2510.06853",
   doi = "10.48550/arXiv.2510.06853"
}

@ARTICLE{bromm13,
   author = "Bromm, Volker",
   title = "Formation of the first stars",
   journal = "Rep. Prog. Phys.",
   year = "2013",
   volume = "76",
   pages = "112901",
   doi = "10.1088/0034-4885/76/11/112901"
}

@ARTICLE{frebel15,
   author = "Frebel, Anna and Norris, John E.",
   title = "Near-Field Cosmology with Extremely Metal-Poor Stars",
   journal = "Annu. Rev. Astron. Astrophys.",
   year = "2015",
   volume = "53",
   pages = "631-688",
   doi = "10.1146/annurev-astro-082214-122423"
}

@ARTICLE{bromm06,
   author = "Bromm, Volker and Loeb, Abraham",
   title = "High-Redshift Gamma-Ray Bursts from Population III Progenitors",
   journal = "Astrophys. J.",
   year = "2006",
   volume = "642",
   pages = "382-388",
   doi = "10.1086/500799"
}

@ARTICLE{heger02,
   author = "Heger, A. and Woosley, S. E.",
   title = "The Nucleosynthetic Signature of Population III",
   journal = "Astrophys. J.",
   year = "2002",
   volume = "567",
   pages = "532-543",
   doi = "10.1086/338487"
}

@ARTICLE{xing23,
   author = "Xing, Qian-Fan and Zhao, Gang and Liu, Zheng-Wei and Heger, Alexander and Han, Zhan-Wen and others",
   title = "A metal-poor star with abundances from a pair-instability supernova",
   journal = "Nature",
   year = "2023",
   volume = "618",
   pages = "712-715",
   doi = "10.1038/s41586-023-06028-1"
}

@ARTICLE{karlsson13,
   author = "Karlsson, Torgny and Bromm, Volker and Bland-Hawthorn, Joss",
   title = "Pregalactic metal enrichment: The chemical signatures of the first stars",
   journal = "Rev. Mod. Phys.",
   year = "2013",
   volume = "85",
   pages = "809-848",
   doi = "10.1103/RevModPhys.85.809"
}

@ARTICLE{aoki14,
   author = "Aoki, W. and Tominaga, N. and Beers, T. C. and Honda, S. and Lee, Y. S.",
   title = "A chemical signature of first-generation very massive stars",
   journal = "Science",
   year = "2014",
   volume = "345",
   pages = "912-915",
   doi = "10.1126/science.1252633"
}

@ARTICLE{beers05,
   author = "Beers, Timothy C. and Christlieb, Norbert",
   title = "The Discovery and Analysis of Very Metal-Poor Stars in the Galaxy",
   journal = "Annu. Rev. Astron. Astrophys.",
   year = "2005",
   volume = "43",
   pages = "531-580",
   doi = "10.1146/annurev.astro.42.053102.134057"
}

@ARTICLE{skuladottir21,
   author = {Sk{\'u}lad{\'o}ttir, {\'A}sa and Salvadori, Stefania and Amarsi, Anish M. and Tolstoy, Eline and Irwin, Michael J. and others},
   title = {Zero-metallicity Hypernova Uncovered by an Ultra-metal-poor Star in the Sculptor Dwarf Spheroidal Galaxy},
   journal = {Astrophys. J. Lett.},
   year = {2021},
   volume = {915},
   pages = {L30},
   doi = {10.3847/2041-8213/ac0dc2}
}

@ARTICLE{hartwig23,
   author = "Hartwig, Tilman and Ishigaki, Miho N. and Kobayashi, Chiaki and Tominaga, Nozomu and Nomoto, Ken'ichi",
   title = "Machine Learning Detects Multiplicity of the First Stars in Stellar Archaeology Data",
   journal = "Astrophys. J.",
   year = "2023",
   volume = "946",
   pages = "20",
   doi = "10.3847/1538-4357/acbcc6"
}

@ARTICLE{ma15,
   author = "Ma, Q. and Maio, U. and Ciardi, B. and Salvaterra, R.",
   title = "PopIII signatures in the spectra of PopII/I GRBs",
   journal = "Mon. Not. R. Astron. Soc.",
   year = "2015",
   volume = "449",
   pages = "3006-3014",
   doi = "10.1093/mnras/stv477"
}

@ARTICLE{inoue07,
   author = "Inoue, Susumu and Omukai, Kazuyuki and Ciardi, Benedetta",
   title = "The radio to infrared emission of very high redshift gamma-ray bursts: probing early star formation through molecular and atomic absorption lines",
   journal = "Mon. Not. R. Astron. Soc.",
   year = "2007",
   volume = "380",
   pages = "1715-1728",
   doi = "10.1111/j.1365-2966.2007.12234.x"
}

@ARTICLE{maiolino24,
   author = "Maiolino, Roberto and {\"U}bler, Hannah and Perna, Michele and Scholtz, Jan and D'Eugenio, Francesco and others",
   title = "JADES. Possible Population III signatures at z = 10.6 in the halo of GN-z11",
   journal = "Astron. Astrophys.",
   year = "2024",
   volume = "687",
   pages = "A67",
   doi = "10.1051/0004-6361/202347087"
}

@ARTICLE{castellano22,
   author = "Castellano, Marco and Fontana, Adriano and Treu, Tommaso and Santini, Paola and Merlin, Emiliano and others",
   title = "Early Results from GLASS-JWST. III. Galaxy Candidates at z  9-15",
   journal = "Astrophys. J. Lett.",
   year = "2022",
   volume = "938",
   pages = "L15",
   doi = "10.3847/2041-8213/ac94d0"
}

@ARTICLE{curtislake23,
   author = "Curtis-Lake, Emma and Carniani, Stefano and Cameron, Alex and Charlot, Stephane and Jakobsen, Peter and others",
   title = "Spectroscopic confirmation of four metal-poor galaxies at z = 10.3-13.2",
   journal = "Nat. Astron.",
   year = "2023",
   volume = "7",
   pages = "622-632",
   doi = "10.1038/s41550-023-01918-w"
}

@ARTICLE{robertson23,
   author = "Robertson, B. E. and Tacchella, S. and Johnson, B. D. and Hainline, K. and Whitler, L. and others",
   title = "Identification and properties of intense star-forming galaxies at redshifts z > 10",
   journal = "Nat. Astron.",
   year = "2023",
   volume = "7",
   pages = "611-621",
   doi = "10.1038/s41550-023-01921-1"
}

@ARTICLE{mesinger11,
   author = "Mesinger, Andrei and Furlanetto, Steven and Cen, Renyue",
   title = "21CMFAST: a fast, seminumerical simulation of the high-redshift 21-cm signal",
   journal = "Mon. Not. R. Astron. Soc.",
   year = "2011",
   volume = "411",
   pages = "955-972",
   doi = "10.1111/j.1365-2966.2010.17731.x"
}

@ARTICLE{mebane18,
   author = "Mebane, Richard H. and Mirocha, Jordan and Furlanetto, Steven R.",
   title = "The Persistence of Population III Star Formation",
   journal = "Mon. Not. R. Astron. Soc.",
   year = "2018",
   volume = "479",
   pages = "4544-4559",
   doi = "10.1093/mnras/sty1833"
}

@ARTICLE{visbal15,
   author = "Visbal, Eli and Haiman, Zoltán and Bryan, Greg L.",
   title = "Limits on Population III star formation in minihaloes implied by Planck",
   journal = "Mon. Not. R. Astron. Soc.",
   year = "2015",
   volume = "453",
   pages = "4456-4466",
   doi = "10.1093/mnras/stv1941"
}

@ARTICLE{singh22,
   author = "Singh, Saurabh and Jishnu, Nambissan T. and Subrahmanyan, Ravi and Udaya Shankar, N. and Girish, B. S. and others",
   title = "On the detection of a cosmic dawn signal in the radio background",
   journal = "Nat. Astron.",
   year = "2022",
   volume = "6",
   pages = "607-617",
   doi = "10.1038/s41550-022-01610-5"
}

@ARTICLE{pochinda24,
   author = "Pochinda, S. and Gessey-Jones, T. and Bevins, H. T. J. and Fialkov, A. and Heimersheim, S. and others",
   title = "Constraining the properties of Population III galaxies with multiwavelength observations",
   journal = "Mon. Not. R. Astron. Soc.",
   year = "2024",
   volume = "531",
   pages = "1113-1132",
   doi = "10.1093/mnras/stae1185"
}

@ARTICLE{hera1,
   author = "HERA Collaboration and Abdurashidova, Zara and Adams, Tyrone and Aguirre, James E. and Alexander, Paul and others",
   title = "Improved Constraints on the 21 cm EoR Power Spectrum and the X-Ray Heating of the IGM with HERA Phase I Observations",
   journal = "Astrophys. J.",
   year = "2023",
   volume = "945",
   pages = "124",
   doi = "10.3847/1538-4357/acaf50"
}

@ARTICLE{abel97,
   author = "Abel, Tom and Anninos, Peter and Zhang, Yu and Norman, Michael L.",
   title = "Modeling primordial gas in numerical cosmology",
   journal = "New Astron.",
   year = "1997",
   volume = "2",
   pages = "181-207",
   doi = "10.1016/S1384-1076(97)00010-9"
}

@ARTICLE{galli98,
   author = "Galli, Daniele and Palla, Francesco",
   title = "The chemistry of the early Universe",
   journal = "Astron. Astrophys.",
   year = "1998",
   volume = "335",
   pages = "403-420",
   doi = "10.48550/arXiv.astro-ph/9803315"
}

@ARTICLE{glover08,
   author = "Glover, S. C. O. and Abel, T.",
   title = "Uncertainties in H2 and HD chemistry and cooling and their role in early structure formation",
   journal = "Mon. Not. R. Astron. Soc.",
   year = "2008",
   volume = "388",
   pages = "1627-1651",
   doi = "10.1111/j.1365-2966.2008.13224.x"
}

@ARTICLE{marigo01,
   author = "Marigo, P. and Girardi, L. and Chiosi, C. and Wood, P. R.",
   title = "Zero-metallicity stars. I. Evolution at constant mass",
   journal = "Astron. Astrophys.",
   year = "2001",
   volume = "371",
   pages = "152-173",
   doi = "10.1051/0004-6361:20010309"
}

@ARTICLE{murphy21,
   author = "Murphy, Laura J. and Groh, Jose H. and Ekström, Sylvia and Meynet, Georges and Pezzotti, Camila and others",
   title = "Grids of stellar models with rotation - V. Models from 1.7 to 120 Msun at zero metallicity",
   journal = "Mon. Not. R. Astron. Soc.",
   year = "2021",
   volume = "501",
   pages = "2745-2763",
   doi = "10.1093/mnras/staa3803"
}

@ARTICLE{hirano14,
   author = "Hirano, Shingo and Hosokawa, Takashi and Yoshida, Naoki and Umeda, Hideyuki and Omukai, Kazuyuki and others",
   title = "One Hundred First Stars: Protostellar Evolution and the Final Masses",
   journal = "Astrophys. J.",
   year = "2014",
   volume = "781",
   pages = "60",
   doi = "10.1088/0004-637X/781/2/60"
}

@ARTICLE{stacy16,
   author = "Stacy, Athena and Bromm, Volker and Lee, Aaron T.",
   title = "Building up the Population III initial mass function from cosmological initial conditions",
   journal = "Mon. Not. R. Astron. Soc.",
   year = "2016",
   volume = "462",
   pages = "1307-1328",
   doi = "10.1093/mnras/stw1728"
}

@ARTICLE{jaura22,
   author = "Jaura, Ondrej and Glover, Simon C. O. and Wollenberg, Katharina M. J. and Klessen, Ralf S. and Geen, Sam and Haemmerlé, Lionel",
   title = "Trapping of H II regions in Population III star formation",
   journal = "Mon. Not. R. Astron. Soc.",
   year = "2022",
   volume = "512",
   pages = "116-136",
   doi = "10.1093/mnras/stac487"
}

@ARTICLE{greif11,
   author = "Greif, Thomas H. and Springel, Volker and White, Simon D. M. and Glover, Simon C. O. and Clark, Paul C. and others",
   title = "Simulations on a Moving Mesh: The Clustered Formation of Population III Protostars",
   journal = "Astrophys. J.",
   year = "2011",
   volume = "737",
   pages = "75",
   doi = "10.1088/0004-637X/737/2/75"
}

@ARTICLE{bromm11,
   author = "Bromm, Volker and Yoshida, Naoki",
   title = "The First Galaxies",
   journal = "Annu. Rev. Astron. Astrophys.",
   year = "2011",
   volume = "49",
   pages = "373-407",
   doi = "10.1146/annurev-astro-081710-102608"
}

@ARTICLE{gao07,
   author = "Gao, L. and Yoshida, N. and Abel, T. and Frenk, C. S. and Jenkins, A. and Springel, V.",
   title = "The first generation of stars in the Lambda cold dark matter cosmology",
   journal = "Mon. Not. R. Astron. Soc.",
   year = "2007",
   volume = "378",
   pages = "449-468",
   doi = "10.1111/j.1365-2966.2007.11814.x"
}

@ARTICLE{oshea15,
   author = "O'Shea, Brian W. and Wise, John H. and Xu, Hao and Norman, Michael L.",
   title = "Probing the Ultraviolet Luminosity Function of the Earliest Galaxies with the Renaissance Simulations",
   journal = "Astrophys. J. Lett.",
   year = "2015",
   volume = "807",
   pages = "L12",
   doi = "10.1088/2041-8205/807/1/L12"
}

@ARTICLE{jaacks18,
   author = "Jaacks, Jason and Thompson, Robert and Finkelstein, Steven L. and Bromm, Volker",
   title = "Baseline metal enrichment from Population III star formation in cosmological volume simulations",
   journal = "Mon. Not. R. Astron. Soc.",
   year = "2018",
   volume = "475",
   pages = "4396-4410",
   doi = "10.1093/mnras/sty062"
}

@ARTICLE{liu20,
   author = "Liu, Boyuan and Bromm, Volker",
   title = "When did Population III star formation end?",
   journal = "Mon. Not. R. Astron. Soc.",
   year = "2020",
   volume = "497",
   pages = "2839-2854",
   doi = "10.1093/mnras/staa2143"
}

@ARTICLE{sarmento22,
   author = "Sarmento, Richard and Scannapieco, Evan",
   title = "The Effects of Radiative Feedback and Supernova-induced Turbulence on Early Galaxies",
   journal = "Astrophys. J.",
   year = "2022",
   volume = "935",
   pages = "174",
   doi = "10.3847/1538-4357/ac815c"
}

@ARTICLE{brauer2025,
   author = "Brauer, Kaley and Emerick, Andrew and Mead, Jennifer and Ji, Alexander P. and Wise, John H. and others",
   title = "AEOS: Star-by-star Cosmological Simulations of Early Chemical Enrichment and Galaxy Formation",
   journal = "Astrophys. J.",
   year = "2025",
   volume = "980",
   pages = "41",
   doi = "10.3847/1538-4357/ada4a1"
}

@ARTICLE{katz25,
   author = "Katz, Harley and Rey, Martin P. and Cadiou, Corentin and Agertz, Oscar and Blaizot, Jeremy and others",
   title = "MEGATRON: Reproducing the Diversity of High-Redshift Galaxy Spectra with Cosmological Radiation Hydrodynamics Simulations",
   journal = "arXiv e-prints",
   year = "2025",
   pages = "arXiv:2510.05201",
   doi = "10.48550/arXiv.2510.05201"
}

@ARTICLE{yoshida08,
   author = "Yoshida, Naoki and Omukai, Kazuyuki and Hernquist, Lars",
   title = "Protostar Formation in the Early Universe",
   journal = "Science",
   year = "2008",
   volume = "321",
   pages = "669",
   doi = "10.1126/science.1160259"
}

@ARTICLE{stacy12,
   author = "Stacy, Athena and Greif, Thomas H. and Bromm, Volker",
   title = "The first stars: mass growth under protostellar feedback",
   journal = "Mon. Not. R. Astron. Soc.",
   year = "2012",
   volume = "422",
   pages = "290-309",
   doi = "10.1111/j.1365-2966.2012.20605.x"
}

@ARTICLE{xu16,
   author = "Xu, Hao and Wise, John H. and Norman, Michael L. and Ahn, Kyungjin and O'Shea, Brian W.",
   title = "Galaxy Properties and UV Escape Fractions during the Epoch of Reionization: Results from the Renaissance Simulations",
   journal = "Astrophys. J.",
   year = "2016",
   volume = "833",
   pages = "84",
   doi = "10.3847/1538-4357/833/1/84"
}

@ARTICLE{zier25,
   author = "Zier, Oliver and Kannan, Rahul and Smith, Aaron and Puchwein, Ewald and Vogelsberger, Mark and others",
   title = "The THESAN-ZOOM project: Population III star formation continues until the end of reionization",
   journal = "Mon. Not. R. Astron. Soc.",
   year = "2025",
   volume = "544",
   pages = "410-429",
   doi = "10.1093/mnras/staf1053"
}

@ARTICLE{grudic22,
   author = "Grudić, Michael Y. and Guszejnov, Dávid and Offner, Stella S. R. and Rosen, Anna L. and Raju, Aman N. and others",
   title = "The dynamics and outcome of star formation with jets, radiation, winds, and supernovae in concert",
   journal = "Mon. Not. R. Astron. Soc.",
   year = "2022",
   volume = "512",
   pages = "216-232",
   doi = "10.1093/mnras/stac526"
}

@ARTICLE{springel03,
   author = "Springel, Volker and Hernquist, Lars",
   title = "Cosmological smoothed particle hydrodynamics simulations: a hybrid multiphase model for star formation",
   journal = "Mon. Not. R. Astron. Soc.",
   year = "2003",
   volume = "339",
   pages = "289-311",
   doi = "10.1046/j.1365-8711.2003.06206.x"
}

@ARTICLE{tegmark97,
   author = "Tegmark, Max and Silk, Joseph and Rees, Martin J. and Blanchard, Alain and Abel, Tom and Palla, Francesco",
   title = "How Small Were the First Cosmological Objects?",
   journal = "Astrophys. J.",
   year = "1997",
   volume = "474",
   pages = "1",
   doi = "10.1086/303434"
}

@ARTICLE{schineider02,
   author = "Schneider, R. and Ferrara, A. and Natarajan, P. and Omukai, K.",
   title = "First Stars, Very Massive Black Holes, and Metals",
   journal = "Astrophys. J.",
   year = "2002",
   volume = "571",
   pages = "30-39",
   doi = "10.1086/339917"
}

@ARTICLE{bromm01,
   author = "Bromm, V. and Ferrara, A. and Coppi, P. S. and Larson, R. B.",
   title = "The fragmentation of pre-enriched primordial objects",
   journal = "Mon. Not. R. Astron. Soc.",
   year = "2001",
   volume = "328",
   pages = "969-976",
   doi = "10.1046/j.1365-8711.2001.04915.x"
}

@ARTICLE{schneider06,
   author = "Schneider, Raffaella and Omukai, Kazuyuki and Inoue, Akio K. and Ferrara, Andrea",
   title = "Fragmentation of star-forming clouds enriched with the first dust",
   journal = "Mon. Not. R. Astron. Soc.",
   year = "2006",
   volume = "369",
   pages = "1437-1444",
   doi = "10.1111/j.1365-2966.2006.10391.x"
}

@ARTICLE{omukai05,
   author = "Omukai, K. and Tsuribe, T. and Schneider, R. and Ferrara, A.",
   title = "Thermal and Fragmentation Properties of Star-forming Clouds in Low-Metallicity Environments",
   journal = "Astrophys. J.",
   year = "2005",
   volume = "626",
   pages = "627-643",
   doi = "10.1086/429955"
}

@ARTICLE{machida06,
   author = "Machida, Masahiro N. and Omukai, Kazuyuki and Matsumoto, Tomoaki and Inutsuka, Shu-ichiro",
   title = "The First Jets in the Universe: Protostellar Jets from the First Stars",
   journal = "Astrophys. J. Lett.",
   year = "2006",
   volume = "647",
   pages = "L1-L4",
   doi = "10.1086/507326"
}

@ARTICLE{vogelsberger14,
   author = "Vogelsberger, Mark and Genel, Shy and Springel, Volker and Torrey, Paul and Sijacki, Debora and others",
   title = "Introducing the Illustris Project: simulating the coevolution of dark and visible matter in the Universe",
   journal = "Mon. Not. R. Astron. Soc.",
   year = "2014",
   volume = "444",
   pages = "1518-1547",
   doi = "10.1093/mnras/stu1536"
}

@ARTICLE{schaye15,
   author = "Schaye, Joop and Crain, Robert A. and Bower, Richard G. and Furlong, Michelle and Schaller, Matthieu and others",
   title = "The EAGLE project: simulating the evolution and assembly of galaxies and their environments",
   journal = "Mon. Not. R. Astron. Soc.",
   year = "2015",
   volume = "446",
   pages = "521-554",
   doi = "10.1093/mnras/stu2058"
}

@ARTICLE{pillepich18,
   author = "Pillepich, Annalisa and Springel, Volker and Nelson, Dylan and Genel, Shy and Naiman, Jill and others",
   title = "Simulating galaxy formation with the IllustrisTNG model",
   journal = "Mon. Not. R. Astron. Soc.",
   year = "2018",
   volume = "473",
   pages = "4077-4106",
   doi = "10.1093/mnras/stx2656"
}

@ARTICLE{deng24,
   author = "Deng, Yunwei and Li, Hui and Liu, Boyuan and Kannan, Rahul and Smith, Aaron and Bryan, Greg L.",
   title = "RIGEL: Simulating dwarf galaxies at solar mass resolution with radiative transfer and feedback from individual massive stars",
   journal = "Astron. Astrophys.",
   year = "2024",
   volume = "691",
   pages = "A231",
   doi = "10.1051/0004-6361/202450699"
}

@ARTICLE{chabrier03,
   author = "Chabrier, Gilles",
   title = "Galactic Stellar and Substellar Initial Mass Function",
   journal = "Publ. Astron. Soc. Pac.",
   year = "2003",
   volume = "115",
   pages = "763-795",
   doi = "10.1086/376392"
}

@ARTICLE{castor75,
   author = "Castor, J. I. and Abbott, D. C. and Klein, R. I.",
   title = "Radiation-driven winds in Of stars.",
   journal = "Astrophys. J.",
   year = "1975",
   volume = "195",
   pages = "157-174",
   doi = "10.1086/153315"
}

@ARTICLE{maraston05,
   author = "Maraston, Claudia",
   title = "Evolutionary population synthesis: models, analysis of the ingredients and application to high-z galaxies",
   journal = "Mon. Not. R. Astron. Soc.",
   year = "2005",
   volume = "362",
   pages = "799-825",
   doi = "10.1111/j.1365-2966.2005.09270.x"
}

@ARTICLE{ekstrom12,
   author = "Ekström, S. and Georgy, C. and Eggenberger, P. and Meynet, G. and Mowlavi, N. and others",
   title = "Grids of stellar models with rotation. I. Models from 0.8 to 120 Msun at solar metallicity (Z = 0.014)",
   journal = "Astron. Astrophys.",
   year = "2012",
   volume = "537",
   pages = "A146",
   doi = "10.1051/0004-6361/201117751"
}

@ARTICLE{li15,
   author = "Li, Miao and Ostriker, Jeremiah P. and Cen, Renyue and Bryan, Greg L. and Naab, Thorsten",
   title = "Supernova Feedback and the Hot Gas Filling Fraction of the Interstellar Medium",
   journal = "Astrophys. J.",
   year = "2015",
   volume = "814",
   pages = "4",
   doi = "10.1088/0004-637X/814/1/4"
}

@ARTICLE{li20,
   author = "Li, Miao and Li, Yuan and Bryan, Greg L. and Ostriker, Eve C. and Quataert, Eliot",
   title = "The Impact of Type Ia Supernovae in Quiescent Galaxies. I. Formation of the Multiphase Interstellar Medium",
   journal = "Astrophys. J.",
   year = "2020",
   volume = "894",
   pages = "44",
   doi = "10.3847/1538-4357/ab86b4"
}

@ARTICLE{nunez17,
   author = "Núñez, Alejandro and Ostriker, Jeremiah P. and Naab, Thorsten and Oser, Ludwig and Hu, Chia-Yu and Choi, Ena",
   title = "Modeling for Stellar Feedback in Galaxy Formation Simulations",
   journal = "Astrophys. J.",
   year = "2017",
   volume = "836",
   pages = "204",
   doi = "10.3847/1538-4357/836/2/204"
}

@ARTICLE{kimm14,
   author = "Kimm, Taysun and Cen, Renyue",
   title = "Escape Fraction of Ionizing Photons during Reionization: Effects due to Supernova Feedback and Runaway OB Stars",
   journal = "Astrophys. J.",
   year = "2014",
   volume = "788",
   pages = "121",
   doi = "10.1088/0004-637X/788/2/121"
}

@ARTICLE{thornton98,
   author = "Thornton, K. and Gaudlitz, M. and Janka, H. -Th. and Steinmetz, M.",
   title = "Energy Input and Mass Redistribution by Supernovae in the Interstellar Medium",
   journal = "Astrophys. J.",
   year = "1998",
   volume = "500",
   pages = "95-119",
   doi = "10.1086/305704"
}

@ARTICLE{chevalier74,
   author = "Chevalier, Roger A.",
   title = "The Evolution of Supernova Remnants. Spherically Symmetric Models",
   journal = "Astrophys. J.",
   year = "1974",
   volume = "188",
   pages = "501-516",
   doi = "10.1086/152740"
}

@ARTICLE{vink2001,
   author = "Vink, Jorick S. and de Koter, A. and Lamers, H. J. G. L. M.",
   title = "Mass-loss predictions for O and B stars as a function of metallicity",
   journal = "Astron. Astrophys.",
   year = "2001",
   volume = "369",
   pages = "574-588",
   doi = "10.1051/0004-6361:20010127"
}

@ARTICLE{hopkins18,
   author = "Hopkins, Philip F. and Wetzel, Andrew and Kereš, Dušan and Faucher-Giguère, Claude-André and Quataert, Eliot and others",
   title = "How to model supernovae in simulations of star and galaxy formation",
   journal = "Mon. Not. R. Astron. Soc.",
   year = "2018",
   volume = "477",
   pages = "1578-1603",
   doi = "10.1093/mnras/sty674"
}

@ARTICLE{novak11,
   author = "Novak, Gregory S. and Ostriker, Jeremiah P. and Ciotti, Luca",
   title = "Feedback from Central Black Holes in Elliptical Galaxies: Two-dimensional Models Compared to One-dimensional Models",
   journal = "Astrophys. J.",
   year = "2011",
   volume = "737",
   pages = "26",
   doi = "10.1088/0004-637X/737/1/26"
}

@ARTICLE{yuan18,
   author = "Yuan, Feng and Yoon, DooSoo and Li, Ya-Ping and Gan, Zhao-Ming and Ho, Luis C. and Guo, Fulai",
   title = "Active Galactic Nucleus Feedback in an Elliptical Galaxy with the Most Updated AGN Physics. I. Low Angular Momentum Case",
   journal = "Astrophys. J.",
   year = "2018",
   volume = "857",
   pages = "121",
   doi = "10.3847/1538-4357/aab8f8"
}

@ARTICLE{zhu23,
   author = "Zhu, Bocheng and Yuan, Feng and Ji, Suoqing and Peng, Yingjie and Ho, Luis C. and others",
   title = "Active galactic nuclei feedback in an elliptical galaxy (III): the impacts and fate of cosmological inflow",
   journal = "Mon. Not. R. Astron. Soc.",
   year = "2023",
   volume = "524",
   pages = "5787-5803",
   doi = "10.1093/mnras/stad2055"
}

@ARTICLE{nomoto13,
   author = "Nomoto, Ken'ichi and Kobayashi, Chiaki and Tominaga, Nozomu",
   title = "Nucleosynthesis in Stars and the Chemical Enrichment of Galaxies",
   journal = "Annu. Rev. Astron. Astrophys.",
   year = "2013",
   volume = "51",
   pages = "457-509",
   doi = "10.1146/annurev-astro-082812-140956"
}

@ARTICLE{kannan14,
   author = "Kannan, R. and Stinson, G. S. and Macciò, A. V. and Hennawi, J. F. and Woods, R. and others",
   title = "Galaxy formation with local photoionization feedback - I. Methods",
   journal = "Mon. Not. R. Astron. Soc.",
   year = "2014",
   volume = "437",
   pages = "2882-2893",
   doi = "10.1093/mnras/stt2098"
}

@ARTICLE{zhu24,
   author = "Zhu, Bocheng and Springel, Volker",
   title = "The effect of local photoionization on the galaxy properties and the circumgalactic medium in simulations of Milky Way-sized galaxies",
   journal = "Mon. Not. R. Astron. Soc.",
   year = "2024",
   volume = "533",
   pages = "4360-4383",
   doi = "10.1093/mnras/stae2047"
}

@ARTICLE{hopkins20,
   author = "Hopkins, Philip F. and Grudić, Michael Y. and Wetzel, Andrew and Kereš, Dušan and Faucher-Giguère, Claude-André and others",
   title = "Radiative stellar feedback in galaxy formation: Methods and physics",
   journal = "Mon. Not. R. Astron. Soc.",
   year = "2020",
   volume = "491",
   pages = "3702-3729",
   doi = "10.1093/mnras/stz3129"
}

@ARTICLE{wg11,
   author = "Wolcott-Green, J. and Haiman, Z. and Bryan, G. L.",
   title = "Photodissociation of H2 in protogalaxies: modelling self-shielding in three-dimensional simulations",
   journal = "Mon. Not. R. Astron. Soc.",
   year = "2011",
   volume = "418",
   pages = "838-852",
   doi = "10.1111/j.1365-2966.2011.19538.x"
}

@ARTICLE{katz96,
   author = "Katz, Neal and Weinberg, David H. and Hernquist, Lars",
   title = "Cosmological Simulations with TreeSPH",
   journal = "Astrophys. J. Suppl.",
   year = "1996",
   volume = "105",
   pages = "19",
   doi = "10.1086/192305"
}

@ARTICLE{kannan25,
   author = "Kannan, Rahul and Puchwein, Ewald and Smith, Aaron and Borrow, Josh and Garaldi, Enrico and others",
   title = "Introducing the THESAN-ZOOM project: radiation-hydrodynamic simulations of high-redshift galaxies with a multi-phase interstellar medium",
   journal = "The Open Journal of Astrophysics",
   year = "2025",
   volume = "8",
   pages = "153",
   doi = "10.33232/001c.145804"
}

@ARTICLE{ferland17,
   author = "Ferland, G. J. and Chatzikos, M. and Guzmán, F. and Lykins, M. L. and van Hoof, P. A. M. and others",
   title = "The 2017 Release Cloudy",
   journal = "Rev. Mex. Astron. Astrofis.",
   year = "2017",
   volume = "53",
   pages = "385-438",
   doi = "10.48550/arXiv.1705.10877"
}

@ARTICLE{truelove97,
   author = "Truelove, J. Kelly and Klein, Richard I. and McKee, Christopher F. and Holliman, II, John H. and Howell, Louis H. and Greenough, Jeffrey A.",
   title = "The Jeans Condition: A New Constraint on Spatial Resolution in Simulations of Isothermal Self-gravitational Hydrodynamics",
   journal = "Astrophys. J. Lett.",
   year = "1997",
   volume = "489",
   pages = "L179-L183",
   doi = "10.1086/310975"
}

@ARTICLE{springel05,
   author = "Springel, Volker and White, Simon D. M. and Jenkins, Adrian and Frenk, Carlos S. and Yoshida, Naoki and others",
   title = "Simulations of the formation, evolution and clustering of galaxies and quasars",
   journal = "Nature",
   year = "2005",
   volume = "435",
   pages = "629-636",
   doi = "10.1038/nature03597"
}

@ARTICLE{angulo12,
   author = "Angulo, R. E. and Springel, V. and White, S. D. M. and Jenkins, A. and Baugh, C. M. and Frenk, C. S.",
   title = "Scaling relations for galaxy clusters in the Millennium-XXL simulation",
   journal = "Mon. Not. R. Astron. Soc.",
   year = "2012",
   volume = "426",
   pages = "2046-2062",
   doi = "10.1111/j.1365-2966.2012.21830.x"
}

@ARTICLE{bromm03firstSN,
   author = "Bromm, Volker and Yoshida, Naoki and Hernquist, Lars",
   title = "The First Supernova Explosions in the Universe",
   journal = "Astrophys. J. Lett.",
   year = "2003",
   volume = "596",
   pages = "L135-L138",
   doi = "10.1086/379359"
}

@ARTICLE{oh2002,
   author = "Oh, S. Peng and Haiman, Zoltán",
   title = "Second-Generation Objects in the Universe: Radiative Cooling and Collapse of Halos with Virial Temperatures above $10^4$ K",
   journal = "Astrophys. J.",
   year = "2002",
   volume = "569",
   pages = "558-572",
   doi = "10.1086/339393"
}

@ARTICLE{kennicutt2012,
   author = "Kennicutt, Robert C. and Evans, Neal J.",
   title = "Star Formation in the Milky Way and Nearby Galaxies",
   journal = "Annu. Rev. Astron. Astrophys.",
   year = "2012",
   volume = "50",
   pages = "531-608",
   doi = "10.1146/annurev-astro-081811-125610"
}

@ARTICLE{eisenstein99,
   author = "Eisenstein, Daniel J. and Hu, Wayne",
   title = "Power Spectra for Cold Dark Matter and Its Variants",
   journal = "Astrophys. J.",
   year = "1999",
   volume = "511",
   pages = "5-15",
   doi = "10.1086/306640"
}

@ARTICLE{planck16,
   author = "Planck Collaboration and Ade, P. A. R. and Aghanim, N. and Arnaud, M. and Ashdown, M. and others",
   title = "Planck 2015 results. XIII. Cosmological parameters",
   journal = "Astron. Astrophys.",
   year = "2016",
   volume = "594",
   pages = "A13",
   doi = "10.1051/0004-6361/201525830"
}

@ARTICLE{escala18,
   author = "Escala, Ivanna and Wetzel, Andrew and Kirby, Evan N. and Hopkins, Philip F. and Ma, Xiangcheng and others",
   title = "Modelling chemical abundance distributions for dwarf galaxies in the Local Group: the impact of turbulent metal diffusion",
   journal = "Mon. Not. R. Astron. Soc.",
   year = "2018",
   volume = "474",
   pages = "2194-2211",
   doi = "10.1093/mnras/stx2858"
}

@ARTICLE{smagorinsky63,
   author = "Smagorinsky, J.",
   title = "General Circulation Experiments with the Primitive Equations",
   journal = "Mon. Weather Rev.",
   year = "1963",
   volume = "91",
   pages = "99",
   doi = "10.1175/1520-0493(1963)091<0099:GCEWTP>2.3.CO;2"
}

@ARTICLE{shen10,
   author = "Shen, S. and Wadsley, J. and Stinson, G.",
   title = "The enrichment of the intergalactic medium with adiabatic feedback - I. Metal cooling and metal diffusion",
   journal = "Mon. Not. R. Astron. Soc.",
   year = "2010",
   volume = "407",
   pages = "1581-1596",
   doi = "10.1111/j.1365-2966.2010.17047.x"
}

@ARTICLE{pakmor2016improving,
   author = "Pakmor, Rüdiger and Springel, Volker and Bauer, Andreas and Mocz, Philip and Munoz, Diego J. and others",
   title = "Improving the convergence properties of the moving-mesh code AREPO",
   journal = "Mon. Not. R. Astron. Soc.",
   year = "2016",
   volume = "455",
   pages = "1134-1143",
   doi = "10.1093/mnras/stv2380"
}

@ARTICLE{eldridge17,
   author = "Eldridge, J. J. and Stanway, E. R. and Xiao, L. and McClelland, L. A. S. and Taylor, G. and others",
   title = "Binary Population and Spectral Synthesis Version 2.1: Construction, Observational Verification, and New Results",
   journal = "Publ. Astron. Soc. Aust.",
   year = "2017",
   volume = "34",
   pages = "e058",
   doi = "10.1017/pasa.2017.51"
}

@ARTICLE{gnedin2009,
   author = "Gnedin, Nickolay Y. and Tassis, Konstantinos and Kravtsov, Andrey V.",
   title = "Modeling Molecular Hydrogen and Star Formation in Cosmological Simulations",
   journal = "Astrophys. J.",
   year = "2009",
   volume = "697",
   pages = "55-67",
   doi = "10.1088/0004-637X/697/1/55"
}

@ARTICLE{hopkins2018fire2,
    author = "Hopkins, Philip F. and Wetzel, Andrew and Kere{\v{s}}, Du{\v{s}}an and Faucher-Gigu{\`e}re, Claude-Andr{\'e} and Quataert, Eliot and Boylan-Kolchin, Michael and Murray, Norman and Hayward, Christopher C. and Garrison-Kimmel, Shea and Hummels, Cameron and Feldmann, Robert and Torrey, Paul and Ma, Xiangcheng and Angl{\'e}s-Alc{\'a}zar, Daniel and Su, Kung-Yi and Orr, Matthew and Schmitz, Denise and Escala, Ivanna and Sanderson, Robyn and Grudi{\'c}, Michael Y. and Hafen, Zachary and Kim, Ji-Hoon and Fitts, Alex and Bullock, James S. and Wheeler, Coral and Chan, T.~K. and Elbert, Oliver D. and Narayanan, Desika",
    title = "FIRE-2 simulations: physics versus numerics in galaxy formation",
    journal = "Mon. Not. R. Astron. Soc.",
    year = "2018",
    volume = "480",
    pages = "800-863",
    doi = "10.1093/mnras/sty1690"
}

@ARTICLE{marinacci2019,
    author = "Marinacci, Federico and Sales, Laura V. and Vogelsberger, Mark and Torrey, Paul and Springel, Volker",
    title = "Simulating the interstellar medium and stellar feedback on a moving mesh: implementation and isolated galaxies",
    journal = "Mon. Not. R. Astron. Soc.",
    year = "2019",
    volume = "489",
    pages = "4233-4260",
    doi = "10.1093/mnras/stz2391"
}

@ARTICLE{genel19,
       author = "Genel, Shy and Bryan, Greg L. and Springel, Volker and Hernquist, Lars and Nelson, Dylan and Pillepich, Annalisa and Weinberger, Rainer and Pakmor, R{\"u}diger and Marinacci, Federico and Vogelsberger, Mark",
        title = "A Quantification of the Butterfly Effect in Cosmological Simulations and Implications for Galaxy Scaling Relations",
      journal = "Astrophys. J.",
         year = 2019,
       volume = "871",
        pages = "21",
          doi = "10.3847/1538-4357/aaf4bb",
}

@ARTICLE{zhu2025,
       author = "Zhu, Bocheng and Springel, Volker",
        title = "Different physical and numerical sources of scatter in the M$_{BH}$-M$_{{\ensuremath{\star}}}$ relation and their connection to galaxy evolution",
      journal = "Mon. Not. R. Astron. Soc.",
         year = 2025,
       volume = "543",
        pages = "2489-2506",
          doi = "10.1093/mnras/staf1559"
}

@BOOK{draine2011,
       author = "Draine, Bruce T.",
        title = "Physics of the Interstellar and Intergalactic Medium",
         year = "2011",
         adsurl = "https://ui.adsabs.harvard.edu/abs/2011piim.book.....D"
         
}

\end{document}